\documentclass[pra,showpacs]{revtex4}
\usepackage{hyperref}
\usepackage{amsmath}
\usepackage{amssymb}
\usepackage{amsthm}
\usepackage{graphics}
\usepackage{graphicx}
\long\def\ca#1\cb{}

\def\CC{{\cal C}}
\def\DC{{\cal D}}

\def\JC{{\cal J}}

\def\SC{{\cal S}}

\def\BC{{\cal B}}
\def\AC{{\cal A}}

\newtheorem{thm1}{Theorem}
\newtheorem{lem1}[thm1]{Lemma}
\newtheorem{lem2}[thm1]{Lemma}

\begin{document}
\title{When a quantum measurement can be implemented locally, and when it cannot}
\author{Scott M. Cohen}
\email{cohensm@duq.edu}
\affiliation{Department of Physics, Duquesne University, Pittsburgh,
Pennsylvania 15282\\
Department of Physics, Carnegie-Mellon University,
Pittsburgh, Pennsylvania 15213}

\begin{abstract}
In the absence of quantum channels, local operations on subsystems and classical communication between parties (LOCC) constitute the most general protocols available on spatially separated quantum systems. Every LOCC protocol implements a separable quantum measurement, but it is known that there exist separable measurements that cannot be implemented by LOCC. A longstanding problem in quantum information theory is to understand the difference between LOCC and the full set of separable measurements. Toward this end, we show in this paper how to construct an LOCC protocol to implement an arbitrary separable measurement whenever such a protocol exists. In addition, given a measurement which cannot be implemented by LOCC within some fixed maximum number of rounds, the method shows explicitly that this is the case.
\end{abstract}

\date{Version of 10 October 2010}

\maketitle
\section{Introduction}
Left to their own devices, quantum systems undergo unitary evolution. They may interact with other quantum systems, but considering all these systems together as a single entity, its state at any given time is related to that at any other time by a unitary transformation. We as scientists, however, often wish to know something about the systems we are studying, so we perform measurements to extract information with the aim of understanding the behavior of these systems. In order to draw conclusions, we must understand the measurements that we make, and perhaps more importantly, we will want to optimize our measurements to maximize the information we can extract given the constraints with which we either choose, or are forced by circumstances, to work. Of course, if we find a measurement that is optimal for one set of circumstances, it may well be that this measurement cannot be performed under other constraints. It is therefore crucial that we have a way to determine when a measurement is possible, and when it is not.

If the system under consideration resides in a single laboratory, then it is purely an experimental question whether or not a given measurement is possible --- do we have the tools and skills, or don't we? If, on the other hand, the system consists of two or more spatially separated subsystems, then it will often be the case that a given measurement simply cannot, \textit{in principle}, be implemented. This question of the ``local implementation" of a measurement is of fundamental interest for our understanding of quantum theory itself, and it also arises naturally in numerous applications considered in the quantum information sciences. Examples of such applications include distributed quantum computing \cite{CiracDistComp}, one-way quantum computing \cite{RaussendorfBriegel}, entanglement distillation \cite{BennettPurify} and manipulation \cite{Nielsen}, local distinguishability of quantum states \cite{Walgate}, local cloning \cite{Anselmi}, and various quantum cryptographic protocols, such as secret sharing \cite{HillerySecret}.

It is not too difficult to describe in words the most general protocol possible for implementing a local measurement. Let us assume there are two subsystems, one (denote it as $A$) located in Alice's laboratory and the other ($B$) in Bob's. One of the parties, say Alice, starts by locally (on $A$) performing a generalized measurement \cite{Kraus} with outcomes corresponding to Kraus operators $A_{i_1}$. If the initial state of the system was $|\Psi_0\rangle$ and Alice obtained outcome $i_1$, the state will now be $(A_{i_1}\otimes I_B)|\Psi_0\rangle$, which is generally no longer normalized, and $I_B$ ($I_A$) is the identity operator on system $B$ ($A$). Alice calls Bob on the telephone and informs him her outcome was $i_1$, after which he performs a measurement on $B$, conditioned on Alice's outcome $i_1$ and described by Kraus operators $B^{(i_1)}_{i_2}$. He then informs Alice that his outcome was $i_2$, after which she performs a measurement with outcomes $A^{(i_1,i_2)}_{i_3}$, and they may continue in this way for an arbitrary number of rounds. From the fact that the probabilities of outcomes obtained at each stage must always sum to unity, one has that for each and every $n$,
\begin{eqnarray}\label{closure}
	I_A = \sum_{i_n}A^{(\SC_n)\dagger}_{i_n}A^{(\SC_n)}_{i_n}\nonumber\\
	I_B = \sum_{i_n}B^{(\SC_n)\dagger}_{i_n}B^{(\SC_n)}_{i_n},
\end{eqnarray}
where $\SC_n$ is a collection of indices $\{i_1,i_2,\cdots,i_{n\!-\!1}\}$ indicating all outcomes obtained in earlier measurements. The final (unnormalized) state of the system is given in terms of these Kraus operators as
\begin{eqnarray}
	|\Psi_f\rangle = \left[(\cdots A^{(i_1,i_2)}_{i_3}A_{i_1})\otimes (\cdots B^{(i_1,i_2,i_3)}_{i_4}B^{(i_1)}_{i_2})\right]|\Psi_0\rangle = \left[\,\widehat A^{(\widehat \SC\,)}\otimes \widehat B^{(\widehat \SC\,)}\right]|\Psi_0\rangle,
\end{eqnarray}
and $\widehat \SC$ denotes the full set of local outcomes (leading to this particular overall final outcome) in the entire process just described.

This process, known as LOCC (for local operations and classical communication), is quite complicated and difficult to analyze in detail. However, we see that the final outcomes are always in terms of product Kraus operators, $\widehat A^{(\widehat \SC\,)}\otimes \widehat B^{(\widehat \SC\,)}$, so we have what is known as a separable measurement \cite{Rains}. Then, if one chooses to focus only on the final outcomes and not on how one actually gets to those outcomes by local measurements, the description is greatly simplified and more easily understood. For this reason, it is quite common to study separable measurements in the hope of gaining a better understanding of LOCC \cite{Gheorghiu2,Anselmi,Chefles,CiracEnt}. It is known, however, that there exist separable measurements that cannot be implemented by LOCC \cite{Bennett9,IBM_CMP,IBM_PRL,Chitambar}.

Therefore before we can truly understand LOCC, we will need to know more about the difference between LOCC and the full set of separable measurements, something about which very little has been known up to the present time. In this paper, we provide an important step toward this goal by showing how to construct an LOCC protocol from an arbitrary separable measurement whenever this is possible. To be precise, by a \textit{separable measurement} we will mean a fixed collection $\{\widehat A_j\otimes \widehat B_j\}$ of distinct product Kraus operators for which there exists a set of positive coefficients, $\{\widehat r_j\}$~\footnote{This definition fixes the outcomes of the measurement, defined by the Kraus operators, but not the probabilities of obtaining those outcomes, which also depend on the $\widehat r_j$.}, such that
\begin{eqnarray}\label{closureAB}
	I_A\otimes I_B = \sum_{j} \widehat r_j\widehat\AC_j\otimes\widehat \BC_j,
\end{eqnarray}
where $\widehat \AC_j=\widehat A_j^\dagger \widehat A_j$ and $\widehat \BC_j=\widehat B_j^\dagger \widehat B_j$ (this definition of a measurement should not be confused with a separable operation, which is more general \cite{superNote}). We emphasize that our definition of a measurement is in terms of the set of Kraus operators, and not just the positive operators \{$\widehat \AC_j\otimes\widehat \BC_j\}$, and that there may be more than one set of coefficients, $\widehat r_j$, such that \eqref{closureAB} is satisfied. Our goal in this paper is then to determine whether or not there exists an LOCC protocol for any one such set of coefficients. 

In the next section, we describe a construction that accomplishes this goal, and then provide a detailed algorithm for this construction. In section~\ref{sct1}, several examples are discussed with the aim of giving the reader a better understanding of how the construction works. Then, in section~\ref{conc}, a summary of the results is given. In appendix~\ref{sct3}, we give a proof that the construction does what we have claimed it does, using two important lemmas, proved in appendices~\ref{sct2} and \ref{ssct1}. Appendix~\ref{sct4} discusses the complexity of the construction.

\section{Main result}\label{main}
Our main result is stated below.

\vspace{.1in}
\textbf{Main Theorem} \textit {Suppose Alice and Bob have a separable measurement they wish to perform. Assuming they restrict themselves to some maximum number of rounds, then the construction described below will allow them to determine whether or not the measurement they have designed can be locally implemented, and if it can, will provide them with the LOCC protocol that does so. Note that this claim applies to an extremely general situation, as it does not matter what task is accomplished by the given measurement.}

\vspace{.1in}
In the next subsection, we describe the construction and explain why it accomplishes what we have just claimed. A detailed algorithm is presented in the subsequent subsection, and a proof of the main theorem can be found in appendix \ref{ssct3a}. 

\subsection{The construction}
First note that \emph{any LOCC protocol can be represented as a tree} consisting of nodes into each of which a single branch enters from another node on its left. Time progresses to the right, and each node at ``level" $n$ is associated with a local Kraus operator, $A^{(\SC_{n})}_{i_{n}}$ or $B^{(\SC_{n})}_{i_{n}}$. The subset of nodes at level $n$ that are all attached via a branch to the same node on their left correspond to a complete measurement, the associated Kraus operators satisfying one or the other of Eqs.~(\ref{closure}). Each node may be associated with the local Kraus operator that is performed at that point in the protocol, or equally well with the ordered product of local Kraus operators that have been performed by that party up to that point. We will find it useful, however, to instead associate to each node the positive operator formed from the latter product by multiplying it on its left by its Hermitian conjugate. That is, to each ($A$) node $\SC_{n\!+\!1}=\{\SC_{n},i_n\}$, we will associate the operator
\begin{eqnarray}
	\AC^{(\SC_n)}_{i_{n}}=A_{i_1}^\dagger A^{(i_1,i_2)\dagger}_{i_3}\cdots A^{(\SC_{n\!-\!2})\dagger}_{i_{n\!-\!2}}A^{(\SC_n)\dagger}_{i_n}A^{(\SC_n)}_{i_n}A^{(\SC_{n\!-\!2})}_{i_{n\!-\!2}}\cdots A^{(i_1,i_2)}_{i_3}A_{i_1}.
\end{eqnarray}
Then, by (\ref{closure}), we have
\begin{eqnarray}\label{ACsum}
	\sum_{i_n}\AC^{(\SC_n)}_{i_{n}} = \AC^{(\SC_{n\!-\!2})}_{i_{n\!-\!2}}.
\end{eqnarray}
As simple as this equation is to obtain, it is nonetheless extremely powerful, as it \emph{must be satisfied at each and every node in an LOCC tree}. What this means is that if we know later parts of an LOCC protocol, we can construct the earlier parts that lead to those later ones. In particular, if we know the final outcomes, we can work backward to attempt to construct a full LOCC protocol. If starting from those final outcomes, we can build every tree that is compatible with (\ref{ACsum}), we can then check whether or not one of those trees corresponds to a complete LOCC protocol.

\begin{figure}
\includegraphics{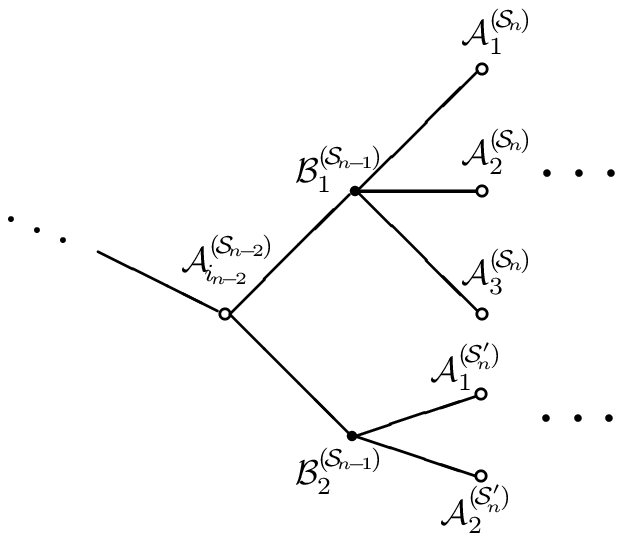}
\caption{\label{fig:treeObs1}Illustration of Eq.~(\ref{ACprop}). The sums, $\AC_1^{(\SC_{n})}+\AC_2^{(\SC_{n})}+\AC_3^{(\SC_{n})}$ and $\AC_1^{(\SC_{n}^\prime)}+\AC_2^{(\SC_{n}^\prime)}$ must be equal to each other and also equal to $\AC_{i_{n\!-\!2}}^{(\SC_{n\!-\!2})}$.}
\end{figure}

Thus, if we sum the positive operators $\AC^{(\SC_n)}_{i_{n}}$ associated with the collection of nodes emerging on the right directly from node $\BC^{(\SC_{{n\!-\!1}})}_{i_{n\!-\!1}}$, we obtain the positive operator associated with the unique node $\AC^{(\SC_{n\!-\!2})}_{i_{n\!-\!2}}$ from which that $\BC^{(\SC_{{n\!-\!1}})}_{i_{n\!-\!1}}$ node emerges. Furthermore, we have the very important observation that this sum is independent of the index $i_{n\!-\!1}$.
This will serve as a useful constraint as there can be many such $B$-nodes emerging from the node $\AC^{(\SC_{n\!-\!2})}_{i_{n\!-\!2}}$, and the sums in (\ref{ACsum}) for all of these $B$-nodes must be the same,
\begin{eqnarray}\label{ACprop}
	\sum_{i_n}\AC^{(\SC_n)}_{i_{n}} = \sum_{i_{\!n}^\prime}\AC^{(\SC_{\!n}^\prime)}_{i_{\!n}^\prime}
\end{eqnarray}
whenever $\SC_{n}$ and $\SC_{\!n}^\prime$ differ only in their last entry, the index $i_{n\!-\!1}$; see Fig.~\ref{fig:treeObs1}. Obviously, there are analogous sums that must be satisfied by the $\BC^{(\SC_{n})}_{i_n}$.

\begin{figure}
\includegraphics{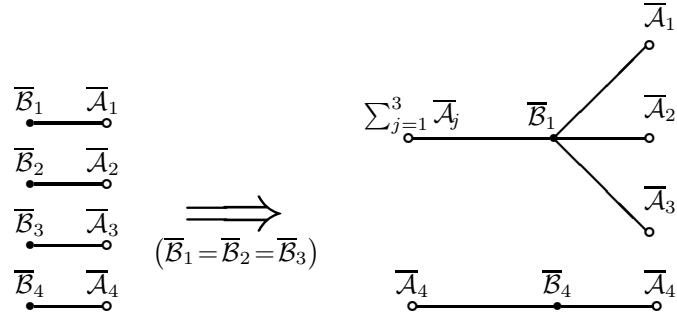}
\caption{\label{fig:treeMerge}Method of construction of an LOCC tree from a set of product operators, $\widehat\AC_j\otimes\widehat\BC_j$. When $\widehat\BC_1=\widehat\BC_2=\widehat\BC_3$, the three $B$-nodes corresponding to these operators can be merged into a single node, after which we attach a new node to its left, labeled by the sum, $\widehat\AC_{1}+\widehat\AC_{2}+\widehat\AC_{3}$.}
\end{figure}

These ideas will now be used to construct a complete LOCC protocol whenever this is possible. Consider first the simplest case where each $\widehat \AC_j\otimes\widehat \BC_j$ appears once and only once in the final set of outcomes of the protocol (the leaves of the tree), the entire collection satisfying (\ref{closureAB}) with $\widehat r_j=1$.  As  illustrated in Fig.~\ref{fig:treeMerge}, start with two-node trees having the operators $\widehat \AC_j$ on the right and the $\widehat \BC_j$ on the left. Find all maximal subsets of the $\widehat \BC_j$ that are equal to each other and merge the corresponding nodes into a single node with multiple branches emerging to the corresponding $A$-nodes. Attach a new $A$-node to the left of each individual (merged) $B$-node, and label these new nodes by the positive operator that is the sum of the $\widehat \AC_j$ that emerge from the given $B$-node (even if there is only a single term in that sum), as shown at the right in Fig.~\ref{fig:treeMerge}, which will insure that (\ref{ACsum}) is always satisfied. By merging multiple nodes into single ones only when all those nodes are equal to each other, we will also insure that the equality is satisfied in (\ref{ACprop}).

This procedure is then iterated: Consider the newest nodes that have just been created at the previous stage, merge subsets of these whose labels are all equal to each other, attach new nodes to the left of these merged nodes, and label each of these newest nodes with a sum of the positive operators that label the nodes that emerge from the corresponding node that was just merged. By induction, all labels will be sums of the $\widehat \AC_j$ or $\widehat \BC_j$ \cite{ACNote}. 
If at some stage all these (sub-)trees merge into a single connected tree, then an LOCC protocol has been identified (readers may find it helpful to study the examples given in section~\ref{sct1}).

If the tree doesn't close, this attempt has failed, but one cannot yet conclude that no LOCC protocol exists for this set of final outcomes. This is because we must first exhaust all possible ways of merging the nodes. For example, if there are three of the $\widehat \BC_1=\widehat \BC_2=\widehat \BC_3$ that are equal to each other, then instead of merging all three into a single node, it may work better to merge only $\widehat \BC_1$ and $\widehat \BC_2$, keeping $\widehat \BC_3$ aside for later use (see Example 4 of section~\ref{sct1}).

Of course, Alice and Bob may also be able to accomplish their task with a protocol that ends with multiple appearances of each of the $\widehat\AC_j\otimes\widehat \BC_j$; allowing such replications introduces additional possible ways of constructing trees. In this case, we must multiply each copy (indexed by $k$) by a positive real factor $\widehat r_{jk}$ to ensure completeness of the measurement [see (\ref{closureAB})]. However, since we will be using (\ref{ACprop}) to compare sums of $\widehat\AC_j$'s and $\widehat \BC_j$'s separately, it will be useful to instead write $\widehat q_{jk}\widehat\AC_j\otimes \widehat p_{jk}\widehat\BC_j$ on the final nodes (see Example 5 in section~\ref{sct1} for more details of why this is useful). That is, by following the above procedure, all $B$-nodes will be labeled by sums of the $\widehat p_{jk}\widehat\BC_j$, and two such nodes can be merged when
\begin{eqnarray}\label{qsums}
	\sum_{j\in \JC}\sum_{k}\widehat p_{jk}\widehat\BC_j=\sum_{j^\prime\in \JC^\prime}\sum_{k^\prime} \widehat p_{j^\prime k^\prime}\widehat\BC_{j^\prime},
\end{eqnarray}
where $\JC$ and $\JC^\prime$ indicate which $\widehat\BC_j$ appear in the sums labeling the two separate nodes in question, and the $k$ ($k^\prime$) sum is determined by which copies of each $\widehat\BC_j$ are present. 

If the positive quantities $\widehat q_{jk},\widehat p_{jk}$ were known, we could simply follow the previous procedure. However, since (a) these factors are unknown, and (b) we do not know how many copies of each $\widehat \AC_j\otimes\widehat \BC_j$ to start with, we are presented with a challenge. Nonetheless, by using the $\widehat q_{jk},\widehat p_{jk}$ as free variables to be determined later by constraints of the form (\ref{qsums}), we will see how to construct an LOCC tree whenever possible, and to thereby determine whether or not an LOCC protocol exists. 

Any positive operator may be thought of as a vector pointing into the ``positive orthant", and positive linear combinations of a set of these vectors lie in the (convex) cone generated by the set. Then, a relationship such as (\ref{qsums}) represents a (non-trivial \cite{trivial}) intersection of the convex cone generated by the $\widehat \BC_j$ for $j\in\JC$ with that for $j\in\JC^\prime$. Any two or more nodes can be identified as satisfying (\ref{qsums}) by looking for a common intersection of their cones \cite{ConvConeIntersect}, and only if their cones intersect can the two nodes be merged. We thus have a way to identify all sets of nodes that can possibly be merged.

One way to construct all possible trees~\footnote{This approach does not actually construct all possible trees, but only those outside of a specific class of exceptions. For a given separable measurement, it is shown in Appendix~\ref{ssct1} that if no LOCC tree exists outside that class, then none exist within that class, either. Therefore, to determine the existence of an LOCC protocol, we only need consider trees outside this class of exceptions.} is as follows: imagine a toolbox initially filled with ``tools" that are two-node trees, $\widehat q_{j1}\widehat\AC_j\otimes \widehat p_{j1}\widehat\BC_j$, as represented at the left of Fig.~\ref{fig:treeMerge}. Identify all intersections of the cones generated by vectors $\widehat\BC_j$; these cones are one-dimensional at this first step. For each intersection, follow the procedure described above of merging nodes (any given tool can be used an arbitrary number of times), adding a new node to the left, and labeling this new node by sums of the associated $\widehat q_{jk}\widehat\AC_j$ (a different $k$ for each use of $\widehat\AC_j\otimes\widehat\BC_j$). Add each of these new trees to the toolbox, while keeping all trees that were already there. For multiple cones that share a common intersection, create all possible new trees, in the same way as was described above for the case that $\widehat\BC_1=\widehat\BC_2=\widehat\BC_3$: one tree for every mutually intersecting subset. In addition, to each new tree in the toolbox, associate a set of constraints of the form (\ref{qsums}), which will be needed at the end to determine the values of the $\widehat p_{jk}$ and $\widehat q_{jk}$. Once this step is completed, we proceed to the next step by looking at all trees presently in the toolbox to identify all intersections of convex cones associated with the left-most (now $A$-)nodes of each of these trees, merge nodes in all possible allowed ways to create new trees, add these to the toolbox, and combine the sets of constraints associated with the merged trees along with the new constraints from the newly merged nodes to obtain a larger set of constraints.

This procedure is then iterated, creating \emph{all}~\footnotemark[\value{footnote}] \emph{possible (connected) trees at each step} that are consistent with (\ref{ACsum}) and (\ref{ACprop}). Continue until one of these trees includes each of the $\widehat\AC_j\otimes \widehat\BC_j$ at least once (we will refer to these as ``complete" trees), there is nothing further one can do, or one has already used the maximum allowed number of rounds (it is necessary to impose a maximum number of rounds because otherwise it appears possible that the procedure could continue indefinitely even when no LOCC exists). If a complete tree is found, we must consider the full set of (linear) constraints that has been associated with that tree, checking that we can find a set of all the $\widehat q_{jk},\widehat p_{jk}\ge0$ that are consistent with these constraints. Note that since in the actual protocol Alice and Bob begin by having done nothing, represented by operators $I_A$ or $I_B$, the tree must close to a ``double-root" with each root labeled by one or the other identity operator, giving two additional constraints. If such a solution can be found, then an LOCC protocol exists. If not, continue the construction to see if another complete tree can be found. Once one has run out of rounds or come to a point that nothing further can be done, then since we know our construction produces a closed tree whenever an LOCC protocol exists, we may conclude that no LOCC protocol exists for this separable measurement in the given maximum number of rounds. For a more detailed proof of this statement, see appendix~\ref{sct3}.

\subsection{Detailed algorithm}\label{algorithm}
We now give a detailed step-by-step algorithm for the construction just described. At various steps along the way in this algorithm, we create multiple copies of certain trees that had previously been constructed. The point of this is to be sure we merge these trees to other ones in all ways it is possible for that tree to be merged. That is, one copy of a given tree is created for each possible way that tree can be merged to other trees, a \emph{different} copy of that particular tree being used for each different merging. 
\begin{enumerate}
  \item \label{enum0} Start with one two-node tree for each of the $\widehat\AC_j\otimes\widehat\BC_j,~j=1,\ldots,N_0$, each with its $B$-node on the left (the choice of $B$-nodes rather than $A$-nodes is arbitrary, as will become clear below), and include positive factors $\widehat q_{j1} (\widehat p_{j1})$ with each $\widehat\AC_j (\widehat\BC_j)$; partition all left-most nodes into equivalence classes, within each of which all nodes are proportional to each other (note that these equivalence classes are entirely unrelated to classes $\CC$ and $\DC$, which are introduced in appendix \ref{sct3}). Set $l=0$, which will serve as a counter for the depth of the trees (this depth is equal to $l+2$).
  \item \label{enum10} WHILE $l<L$
  \item \label{enum1} Increment $l$. For each equivalence class and for each subset in that class that contains at least one member: (a) CREATE one copy of each tree in the given subset; (b) MERGE left-most nodes of all trees within the subset into a single node and relabel the coefficients $\widehat q_{jk},\widehat p_{jk}$ with a unique value of the $k$ index for every different appearance of $\widehat\AC_j\otimes\widehat\BC_j$ at the leaves within the combined tree, adjusting those $k$ indices throughout the rest of the tree to be consistent [according to \eqref{ACsum}] with its leaf labels; (c) EXTEND the tree by attaching a new node to the left, and label that new node to obey \eqref{ACsum} (note that this must also be done for each one-tree subset, as is illustrated for $\widehat\AC_4\otimes\widehat\BC_4$ at the bottom right of Figure~2 --- this is why it does not matter that we chose to start with all the $B$-nodes on the left, since we will next do the same thing with all the $A$-nodes on the left); (d) RECORD all constraints that the merged nodes must be equal (at the first step, these will be of the form $\widehat p_{ik}\widehat\BC_i=\widehat p_{jk^\prime}\widehat\BC_j$ when node $\widehat\BC_i$ is merged with node $\widehat\BC_j$ and are constraints on the $\widehat p$'s). Number the \emph{new} trees sequentially from $N_{l-1}+1$ to $N_l$, where $N_l$ is the total number of trees at this stage (including trees of all depths constructed so far), which does not exceed $2^{N_{l-1}}-1$. [Note that we can ignore each equivalence class that is identical to one that was previously present at an earlier pass through this algorithm, as all trees that can be constructed from that equivalence class have already been constructed.]
  \item \label{enum2} FOR $m=N_{l-1}+1$ to $N_l$

  \item \label{enum4} IF the $m^{\textrm{th}}$ tree includes each of the $\widehat\AC_j\otimes\widehat\BC_j$ at least once, then: For the collection of all constraints recorded in (multiple passes through) step \ref{enum1} for the $m^{\textrm{th}}$ tree, check to see if there exists a solution for the $\widehat p_{jk},\widehat q_{jk}$, including constraints that the two left-most nodes in the tree under consideration are labeled by $I_A$ and $I_B$ (we start this FOR loop at $N_{l-1}+1$ since all the earlier trees have already been examined; for $l=1$, we know that the first $N_0$ of the trees have only a single $\widehat\AC_j\otimes\widehat\BC_j$, so cannot include each of them at least once). If such a solution exists, we are done, having identified an LOCC protocol, so exit and END this algorithm. 
    \item \label{enum44} END FOR ($m$)
\item \label{enum3}We now have an expanded set of trees all having the same `type' ($A$ or $B$) of nodes on the left, labeled by sums of $\widehat q_{jk}\widehat\AC_j$ or $\widehat p_{jk}\widehat\BC_j$. Consider the convex cones generated by the sets $\{\widehat\AC_j\}$ or $\{\widehat\BC_j\}$ appearing in each such sum for the left-most nodes, and identify an equivalence class for each subset of these nodes such that the associated convex cones are mutually intersecting (a given tree will generally be included in multiple equivalence classes). Each such intersection implies that the associated trees can be merged, which we will do next, so go back to step \ref{enum10} and repeat. However, we only need to look for \emph{new} equivalence classes, involving newly constructed trees (along with all previous ones), since we've already constructed all trees that derive from the other equivalence classes. If there are no new classes then no new trees can be constructed, which means since no previously constructed tree has been found to be LOCC, then no LOCC protocol exists for this measurement, no matter how many rounds are allowed. Therefore, exit and END this algorithm.
  \item END WHILE ($l$)
\end{enumerate}
Note that as we loop through the WHILE loop, we keep \emph{all} trees for the next round, including not just those constructed in the present round, but also those from all previous rounds (we also keep all constraints). This makes it possible for multiple trees of differing size and structure and such that several of them each include the same $\widehat\AC_j\otimes\widehat\BC_j$ (or even several $\widehat\AC_j\otimes\widehat\BC_j$ that are repeated in this way), to be merged together into a single tree. Example 4 of the next section illustrates in a detailed way how the algorithm works.

\section{Examples}\label{sct1}
Here we provide a few additional illustrative examples to make more concrete our method of construction of an LOCC protocol from a set of product operators, as presented in the previous section.
\begin{description}
  \item[Example 1] Let us begin with a simple example, for which it will be easy to see that no LOCC protocol exists. Suppose we have a separable measurement with a corresponding set of positive operators, $\widehat \AC_j\otimes\widehat \BC_j$, for which no two of the $\widehat \AC_j$ (and no two of the $\widehat \BC_j$) are equal to each other (technically, we should require that no two are proportional to each other, as we can always remove a positive factor from an $\widehat\AC_j$ and place it on the $\widehat\BC_j$ without changing $\widehat\AC_j\otimes\widehat\BC_j$, but we can assume that the $\widehat\AC_j$ are all normalized). Since no two of the $\widehat \AC_j$ are equal, there is no possibility that any two (or more) of the $\widehat \BC_j$ can emerge from the same $A$-node, and vice-versa. Hence, it is not possible to merge any of the original two-node trees that we start out with, and which represent the outcomes of our desired measurement, implying that there is no way to even begin to construct an LOCC tree.
  \item[Example 2] Here is another simple example, this time one for which an LOCC does exist, where Alice and Bob each do complete projective measurements on their local systems. Write $\widehat \AC_{j_1}\otimes\widehat \BC_{j_2}=[j_1]_A[j_2]_B$, where $[j_1]_{A}=|j_1\rangle_{A}\langle j_1|$ and similarly for $[j_2]_B$. For each projector $[j_2]_B,~(j_2=0,\cdots,d_B-1),$ onto one of Bob's standard basis states, $j_1$ runs through all values $0,\cdots,d_A-1$, so $\{\widehat \AC_{j_1}\otimes\widehat \BC_{j_2}\}$ is a set of projectors onto a complete basis of Alice and Bob's full Hilbert space. Then, at the first step in our construction and for each value of $j_1$, connect the $A$-nodes of all the two-node trees that are labeled by this $[j_1]_A$. Then, the $B$-nodes that emerge to the right of each of these merged $A$-nodes sum to the identity operator. The new $B$-nodes that we attach to the left of the merged $A$-nodes will then all be labeled by $I_B$. Since these new $B$-nodes all have labels that are equal to each other, they can all be merged into a single node from which emerge (to their right) nodes that constitute a complete set of projectors, $[j_1]_A$, which add to $I_A$. Hence, the new $A$-node that we now attach to the left of the merged $I_B$-node will be labeled by this sum, that is, by $I_A$, so we have a double-rooted tree with the two roots each labeled by one of the identity operators, as required for an LOCC protocol.

Note that precisely this same construction continues to work if we replace $[j_2]_B$ by $[\phi^{(j_1)}_{j_2}]=|\phi^{(j_1)}_{j_2}\rangle_B\langle\phi^{(j_1)}_{j_2}|$, with each set $\{|\phi^{(j_1)}_{j_2}\rangle_B\}$ being a complete orthonormal basis, so that Bob does a different projective measurement for each of the outcomes $j_1$ of Alice's measurement. The reader may find it useful to work this example out by drawing the pictures and explicitly constructing the tree, say for the case that both parties hold qubits and Bob measures either in the $\sigma_z$ or $\sigma_x$ basis depending on Alice's outcome.
  \item[Example 3]Now a more involved example, where I demonstrate for a $3\times3$ system that the separable measurement consisting of projectors onto the $9$ states of \cite{Bennett9} cannot be implemented by LOCC. These states were the first example of what has been called ``nonlocality without entanglement", and the projectors onto them are given by
\begin{eqnarray}\label{B9}
\widehat\AC_1\otimes\widehat\BC_1 & = & [1]_A[1]_B,\nonumber\\
\widehat\AC_2\otimes\widehat\BC_2 & = & [0]_A[0+1]_B,\nonumber\\
\widehat\AC_3\otimes\widehat\BC_3 & = & [0]_A[0-1]_B,\nonumber\\
\widehat\AC_4\otimes\widehat\BC_4 & = & [2]_A[1+2]_B,\nonumber\\
\widehat\AC_5\otimes\widehat\BC_5 & = & [2]_A[1-2]_B,\nonumber\\
\widehat\AC_6\otimes\widehat\BC_6 & = & [1+2]_A[0]_B,\nonumber\\
\widehat\AC_7\otimes\widehat\BC_7 & = & [1-2]_A[0]_B,\nonumber\\
\widehat\AC_8\otimes\widehat\BC_8 & = & [0+1]_A[2]_B,\nonumber\\
\widehat\AC_9\otimes\widehat\BC_9 & = & [0-1]_A[2]_B,
\end{eqnarray}
where, for example, $[0+1]=(|0\rangle+|1\rangle)(\langle0|+\langle1|)/2$. These are the positive operators which we will use to label the initial two-node trees at the beginning of our construction.

Starting with Bob's nodes, we see that $\widehat\BC_6=\widehat\BC_7$ and $\widehat\BC_8=\widehat\BC_9$, so we may merge each of these pairs of $B$-nodes (and no others) into single nodes. Then, since $\widehat\AC_6+\widehat\AC_7=[1]_A+[2]_A$ and $\widehat\AC_8+\widehat\AC_9=[0]_A+[1]_A$, we have the new three-level trees shown in Fig.~\ref{fig:treeB9Step1}.
\begin{figure}
\includegraphics{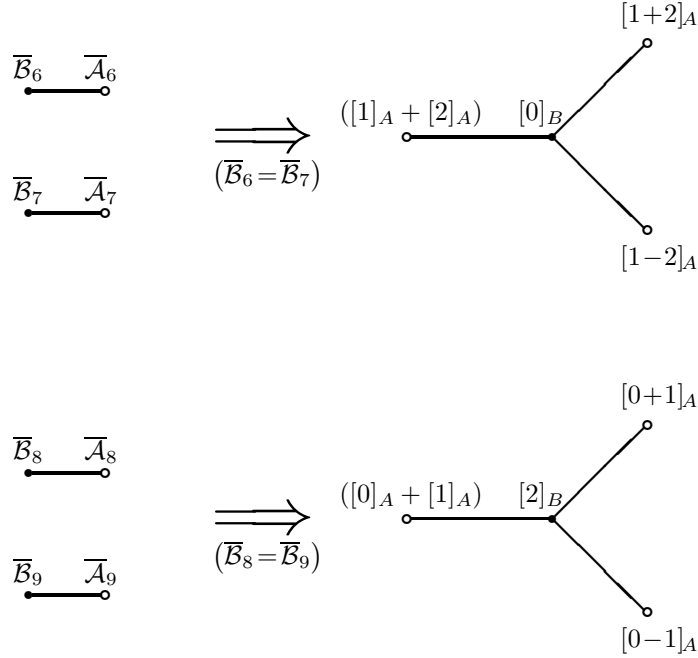}
\caption{\label{fig:treeB9Step1}First step of our attempt at constructing an LOCC tree from projectors onto the states of \cite{Bennett9}, see Eq.~(\ref{B9}). Merging of the two pairs of $B$-nodes, $\widehat\BC_6=\widehat\BC_7=[0]_B$ and $\widehat\BC_8=\widehat\BC_9=[2]_B$, is shown.}
\end{figure}
We next look at these new $A$-nodes together with all the original ones for $j=1,\ldots,9$, and notice that again there are only two pairs that can be merged, $\widehat\AC_2=\widehat\AC_3$ and $\widehat\AC_4=\widehat\AC_5$. Merging these looks very similar to what was just done in Fig.~\ref{fig:treeB9Step1}. Following this, we return to the $B$-nodes to discover that no two labels are equal to each other, as is also the case for the $A$-nodes. There is nothing more that can be done from this point, and since this is the only way to begin building a tree, we see that no four-level trees can be constructed for the operators of \eqref{B9} (other than by trivially adding new nodes to the left of the existing trees, as is illustrated for $\widehat\AC_4\otimes\widehat\BC_4$ at the bottom right of Figure~2, but without additional merging of any trees together). Therefore, there is no LOCC tree compatible with this set of measurement operators, and by the argument in the previous section that every LOCC protocol corresponds to an LOCC tree, we may thus conclude that no LOCC protocol exists for this separable measurement.  In fact, it is clear that even if we allow each of the local projectors in this measurement to be varied slightly (that is, by replacing each $\widehat\AC_j\otimes\widehat\BC_j$ by another operator that differs from it by a small amount), there will still be nothing one can do after these two steps of merging nodes (that is, if one can even still do those two steps). Hence, this separable measurement cannot even be closely approximated by LOCC. Note also that this conclusion holds no matter what the Kraus operators $\widehat A_j\otimes\widehat B_j$ happen to be, so long as they correspond to the positive operators of (\ref{B9}), generalizing the discussion around Eqs.~(59)-(61) in section VII of \cite{Bennett9}.
  \item [Example 4]
Let us now give an example to illustrate how the algorithm of section~\ref{algorithm} works. First, we will describe in general terms the example and how to construct an LOCC protocol, and then we will go back and show how to do this by using the algorithm directly.

Consider a set of five product operators $\{\widehat \AC_{j}\otimes\widehat \BC_{j}\}$ satisfying the constraints
\begin{eqnarray}\label{exmpl}
	\widehat \BC_1&=&\widehat \BC_2=\widehat \BC_3\nonumber\\
	\widehat \BC_5&=&\widehat \BC_1+\widehat \BC_4\nonumber\\
	I_B&=&\widehat \BC_3+\widehat \BC_5\nonumber\\
	\widehat \AC_4&=&\widehat \AC_1+\widehat \AC_2\nonumber\\
	I_A&=&\widehat \AC_3=\widehat \AC_4+\widehat \AC_5,
\end{eqnarray}
and finally that there are no other linear constraints satisfied by these operators. Since no two of the $\widehat \AC_j$'s are proportional to each other, we cannot merge any of the $A$-nodes to begin the construction. We can, on the other hand, merge the three nodes $\widehat \BC_1$, $\widehat \BC_2$, and $\widehat \BC_3$ as was done in Fig.~2. If we start with this step, however, a bit of thought will lead one to conclude that this leaves a situation where there is nothing else that can be done. No LOCC protocol can be generated by starting the construction this way. 

Nonetheless, there is an LOCC protocol for this measurement, as can be seen if we start the construction by merging only the nodes $\widehat \BC_1$ and $\widehat \BC_2$ and labeling this merged node by $\widehat\BC_1$. Then, to the left of this merged node, add a new $A$-node labeled by $\widehat \AC_1+\widehat \AC_2=\widehat \AC_4$, and then merge this new node with node $\widehat \AC_4$. Add a new $B$-node labeled as $\widehat \BC_1+\widehat \BC_4=\widehat \BC_5$ to the left of this and merge it with node $\widehat \BC_5$. Finally, add a new $A$-node labeled as $\widehat \AC_4+\widehat \AC_5$ to the left of this and merge it with node $\widehat \AC_3$. The reader may wish to verify that this construction leads to the tree shown in the upper part (a) of Fig.~\ref{fig:treeFinal}. In the lower part (b) of this figure, we have the same tree but now labeled by the quantities $\AC_{i_n}^{(\SC_n)}$ and $\BC_{i_n}^{(\SC_n)}$ introduced in the preceding section. By comparing these two trees, the latter quantities can be identified as sums of the $\widehat\AC_j$ and $\widehat\BC_j$, respectively, after which it is easy to see that Eqs.~\eqref{ACsum} and \eqref{ACprop} are satisfied (as is guaranteed by our construction).
\begin{figure}
\includegraphics{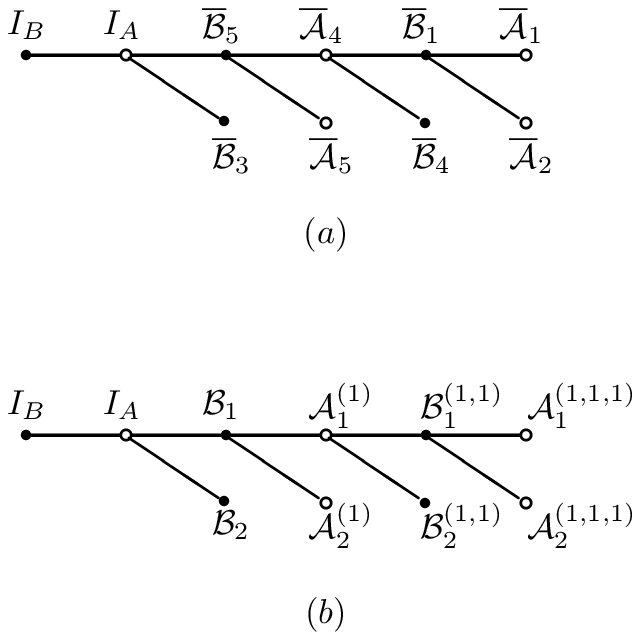}
\caption{\label{fig:treeFinal}The tree constructed for the example of Eq. \eqref{exmpl}. In (a), the nodes are labeled by the desired measurement operators, $\widehat\AC_j$ and $\widehat\BC_j$; while in (b) the nodes are labeled by $\AC^{(\SC_n)}_{i_n}$ and $\BC^{(\SC_n)}_{i_n}$, the quantities appearing in Eqs.~\eqref{ACsum} and \eqref{ACprop}. By comparing these two trees, the latter quantities can be identified in terms of the former.}
\end{figure}

To illustrate directly how our algorithm works, let us reconsider this example. We start with five two-level trees, $B$-nodes on the left as shown in part (a) of figure \ref{fig:treeAlgorithm}, one for each of the $\widehat\AC_j\otimes\widehat\BC_j$. In step \ref{enum0} of the algorithm, identify three equivalence classes, $\mathit{E}_1=\{\widehat\BC_1,\widehat\BC_2,\widehat\BC_3\}$, $\mathit{E}_2=\{\widehat\BC_4\}$, and $\mathit{E}_3=\{\widehat\BC_5\}$. For step \ref{enum1}(a), since $\mathit{E}_2$ and $\mathit{E}_3$ have only a single subset each, creating a copy really means simply keeping that tree; however, for $\mathit{E}_1$ we have seven distinct subsets: $\mathit{s}_{11}=\{\widehat\BC_1\}$, $\mathit{s}_{12}=\{\widehat\BC_2\}$, $\mathit{s}_{13}=\{\widehat\BC_3\}$, $\mathit{s}_{14}=\{\widehat\BC_1,\widehat\BC_2\}$, $\mathit{s}_{15}=\{\widehat\BC_1,\widehat\BC_3\}$, $\mathit{s}_{16}=\{\widehat\BC_2,\widehat\BC_3\}$, $\mathit{s}_{17}=\{\widehat\BC_1,\widehat\BC_2,\widehat\BC_3\}$. The first three subsets give back the original trees for those $B$-nodes, and then we create additional copies of each of these for each appearance in one of the other subsets. Then for each of those latter subsets we: [step \ref{enum1}(b)] merge the corresponding $B$-nodes, [step \ref{enum1}(c)] extend by adding an additional $A$-node, and label the latter by a sum of the $\widehat q_{jk}\widehat\AC_j$ that appear in that tree, and [step \ref{enum1}(d)] record the constraints (which will be similar to the first line of \eqref{exmpl}, but with added factors $\widehat p_{jk}$). The result is shown in part (b) of figure \ref{fig:treeAlgorithm} (though the $\widehat q_{jk},\widehat p_{jk}$ have been omitted to avoid too much clutter in the figure).

For steps \ref{enum2} to \ref{enum44} we look through these trees that were just constructed and see that none of them contains each of the $\widehat\AC_j\otimes\widehat\BC_j$ at least once. Therefore, we continue to step \ref{enum3} to identify new equivalence classes from intersections of convex cones associated with the sums of $\widehat\AC_j$ labeling the left-most nodes in the trees in part (b) of figure \ref{fig:treeAlgorithm}. The only such equivalence class that contains more than one $A$-node in it is $\mathit{E}_4=\{\widehat q_{4k}\widehat \AC_4,\widehat q_{1k}\widehat \AC_1+\widehat q_{2k}\widehat \AC_2\}$, and the only new tree that is constructed from this class is shown in part (c) of figure \ref{fig:treeAlgorithm}. The remainder of the construction proceeds in the same way, eventually yielding the LOCC tree shown in figure \ref{fig:treeFinal}.
\begin{figure}
\includegraphics{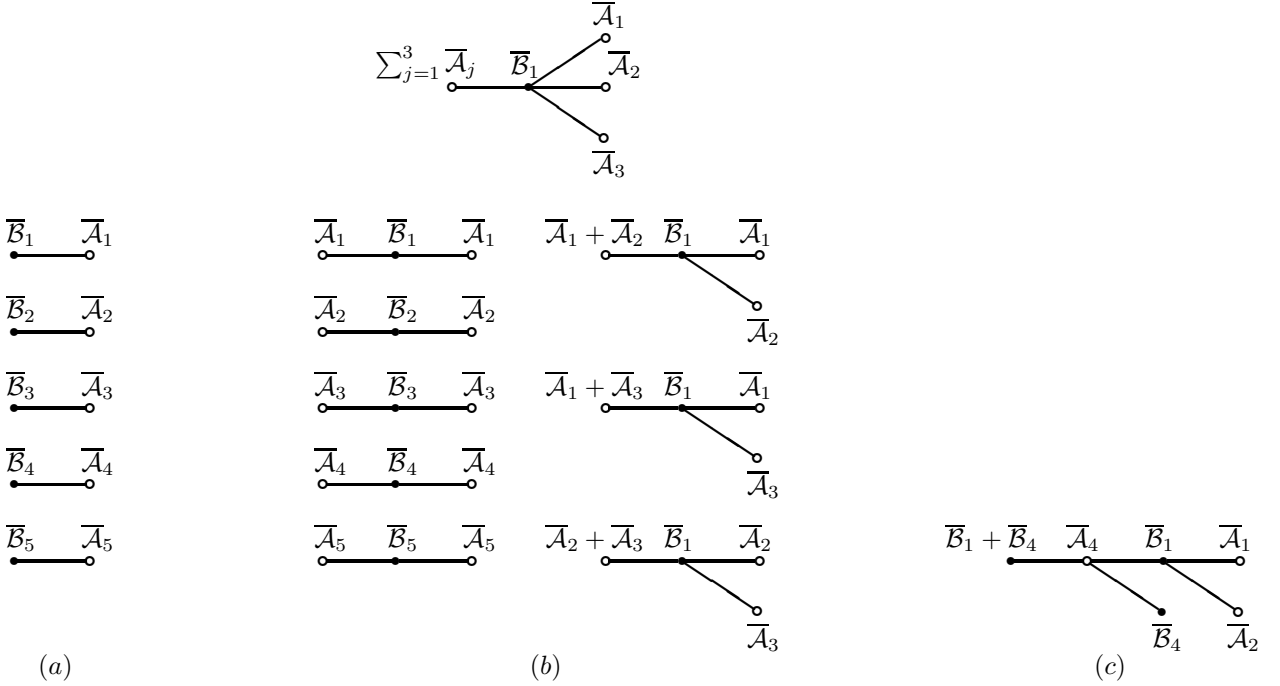}
\caption{\label{fig:treeAlgorithm}The trees constructed by the first two passes through our algorithm for the example of Eq.~\eqref{exmpl}: (a) the five two-level trees representing the five $\widehat\AC_j\otimes\widehat\BC_j$, which are the input to the algorithm; (b) after the first pass through the algorithm, we keep all the original two-level trees, now with one additional node added to turn them into three-level trees, and add four more trees corresponding to the four different ways to merge the three proportional $B$-nodes, $\widehat\BC_1,\widehat\BC_2,\widehat\BC_3$; (c) after the second pass through the algorithm we still have all three-level trees that are shown in (b), but with another added node (these trees are not shown here), along with one additional tree constructed from merging two trees from (b), the one that had $\widehat \AC_4$ as its left-most node and the one that had $\widehat \AC_1+\widehat \AC_2$ as its left-most node. [The reader should mentally insert the factors $\widehat q_{jk},\widehat p_{jk}$, which have been omitted for clarity, to avoid cluttering up the figure.]}
\end{figure}

  \item [Example 5 $\{$Why the $\widehat q_{jk},\widehat p_{jk}$ are useful$\}$] 
Since we want to allow for each $\widehat\AC_j\otimes\widehat\BC_j$ to appear on multiple different leaves, it is necessary to introduce positive factors $\widehat r_{jk}$ multiplying these operators, with $k$ labeling the different appearances of $\widehat\AC_j\otimes\widehat\BC_j$, in order that the sum over all leaves will equal $I_A\otimes I_B$, the required completeness condition on the overall measurement. That is, with $\widehat r_j=\sum_k\widehat r_{jk}$,
\begin{equation}
I_A\otimes I_B=\sum_j\widehat r_j\widehat\AC_j\otimes\widehat\BC_j,
\end{equation}
which is (3). However, since the conditions \eqref{ACsum} and \eqref{ACprop} are conditions on the $\widehat\AC_j$ or $\widehat\BC_j$ separately, it will be useful to write $\widehat r_{jk}=\widehat q_{jk}\widehat p_{jk}$ and consider product operators $\widehat q_{jk}\widehat\AC_j\otimes \widehat p_{jk}\widehat\BC_j$ instead of just $\widehat r_{jk}\widehat\AC_j\otimes \widehat\BC_j$. Let us illustrate why this is useful, and then we will give a complete example.

Consider three of the operators, $\widehat\AC_j\otimes\widehat\BC_j,~j=1,2,3$, and suppose $\widehat\BC_1\ne\widehat\BC_2$ but $\widehat r_{11}\widehat\BC_1=\widehat r_{21}\widehat\BC_2$ for some positive constants $\widehat r_{11}$ and $\widehat r_{21}$. Then, we cannot merge $\widehat\AC_1\otimes\widehat\BC_1$ and $\widehat\AC_2\otimes\widehat\BC_2$, but can merge $\widehat r_{11}\widehat\AC_1\otimes\widehat\BC_1$ with $\widehat r_{21}\widehat\AC_2\otimes\widehat\BC_2$ at the $B$-nodes, labeling this merged node as $\widehat r_{11}\widehat\BC_1$. Following our method of construction, after merging these two $B$-nodes, we introduce a new node to the left and label it by the sum of the $A$-nodes that emerge to the right of these two merged $B$-nodes; that is, by $\widehat\AC_1+\widehat\AC_2$. Now, it may be that this sum is not equal to $\widehat\AC_3$, so the $\widehat\AC_3$ node cannot be merged into the tree we've just constructed. However, if $\widehat q_{31}\widehat\AC_3=\widehat q_{11}\widehat\AC_1+\widehat q_{21}\widehat\AC_2$ for some positive constants $\widehat q_{11}$, $\widehat q_{21}$, and $\widehat q_{31}$, then we should include a tree that has $\widehat q_{31}\widehat\AC_3\otimes \widehat p_{31}\widehat\BC_3$ merged at the $A$-node with a tree created by merging $\widehat q_{11}\widehat\AC_1\otimes \widehat p_{11}\widehat\BC_1$ with $\widehat q_{21}\widehat\AC_2\otimes \widehat p_{21}\widehat\BC_2$ at the $B$-node. This can be done by using $\widehat q_{jk}$ and $\widehat p_{jk}$ instead of the single coefficient $\widehat r_{jk}$, leading to the two constraints,
\begin{align}
\widehat p_{11}\widehat\BC_1&=\widehat p_{21}\widehat\BC_2\notag\\
\widehat q_{31}\widehat\AC_3&=\widehat q_{11}\widehat\AC_1+\widehat q_{21}\widehat\AC_2,
\end{align}
and along with all other constraints, these will determine the set of coefficients $\widehat q_{jk},\widehat p_{jk}$.

Now, the complete example to demonstrate these ideas. This example will, in addition, show how one of the $\widehat\AC_j\otimes\widehat\BC_j$ is used more than once, and we will also explicitly write down the constraints obtained as we merge nodes together. 

Consider a measurement having seven product operators, for which the local positive operators satisfy the following conditions.
\begin{eqnarray}\label{eqnEx5}
	\widehat \BC_1&=&2\widehat \BC_2=3\widehat \BC_3\nonumber\\
	\widehat \BC_6&=&\widehat \BC_1+\widehat \BC_4\nonumber\\
	\widehat \BC_7&=&\widehat \BC_1+2\widehat \BC_5\nonumber\\
	I_B&=&\widehat \BC_6+\widehat \BC_7\nonumber\\
	2\widehat \AC_4&=&\widehat \AC_1+\widehat \AC_2\nonumber\\
	3\widehat \AC_5&=&\widehat \AC_1+\widehat \AC_3\nonumber\\
	I_A&=&\widehat \AC_6+2\widehat \AC_4=\widehat \AC_7+3\widehat \AC_5.
\end{eqnarray}
It is not necessary for us to have discovered these conditions ahead of time, only that we know the operators themselves and are able to check for when positive linear combinations of some of them can be equal to positive linear combinations of others (intersections of convex cones); in the following we will assume that these conditions are not yet known.

Begin by checking for sets of the $\widehat\BC_j$ that are proportional to each other. We find that the only such cases are for $\widehat\BC_1,\widehat\BC_2,\widehat\BC_3$, which are all proportional. We could merge all three together or any of the three pairs, and in general we will want to check all these possibilities, but because of \eqref{eqnEx5} we will see below that only the pairs $j=1,2$ and $j=1,3$ are needed. As we are using $j=1$ twice, multiply each use of $\widehat\AC_1$ and $\widehat\BC_1$ by different coefficients $\widehat q_{11},\widehat p_{11}$ for the first use and $\widehat q_{12},\widehat p_{12}$ for the second one. Then, merging these pairs of $B$-nodes leads to the constraints,
\begin{eqnarray}\label{eqn1}
	\widehat p_{11}\widehat \BC_1&=&\widehat p_2\widehat \BC_2\nonumber\\
	\widehat p_{12}\widehat \BC_1&=&\widehat p_3\widehat \BC_3
\end{eqnarray}
(one could replace $p_2$ by $p_{21}$ and similarly for $p_3$, but since we will end up using only one copy of each of these operators in building a tree, there will be no need for the extra index in this particular example). Attach a new node to the left of each of these merged trees, one of which will be labeled by $\widehat q_{11}\widehat \AC_1+\widehat q_{2}\widehat \AC_2$ and the other by $\widehat q_{12}\widehat \AC_1+\widehat q_{3}\widehat \AC_3$.

We next check the $A$-nodes to see which ones can be merged. By \eqref{eqnEx5}, we will find that $\widehat \AC_4$ lies within the convex cone formed by $\widehat \AC_1$ and $\widehat \AC_2$, while $\widehat \AC_5$ lies within the convex cone formed by $\widehat \AC_1$ and $\widehat \AC_3$, so we can merge the $\widehat \AC_4$ node and the $\widehat \AC_5$ node into the appropriate one of the trees created at the previous step where $B$-nodes were merged to start things off. Doing this and attaching new nodes on the left leaves us with part (a) of Figure~\ref{fig:treeEx5}. The constraints obtained for these mergings are
\begin{eqnarray}\label{eqn2}
	\widehat q_4\widehat \AC_4&=&\widehat q_{11}\widehat \AC_1+\widehat q_2\widehat \AC_2\nonumber\\
	\widehat q_5\widehat \AC_5&=&\widehat q_{12}\widehat \AC_1+\widehat q_3\widehat \AC_3.
\end{eqnarray}
\begin{figure}
\includegraphics{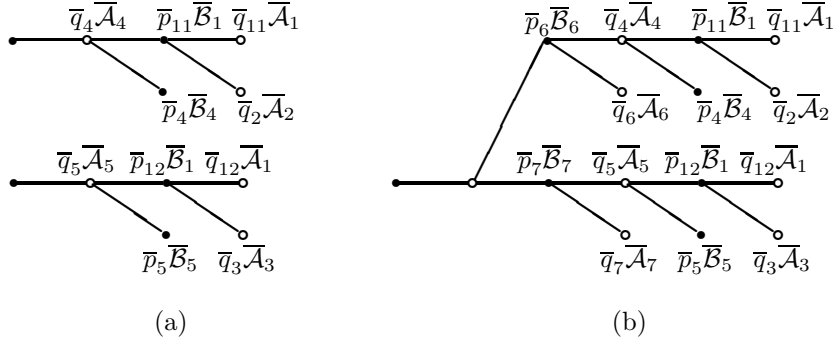}
\caption{\label{fig:treeEx5}The tree constructed for Example 5: (a) mid-way through the construction; (b) the final tree.}
\end{figure}
The left-most nodes of the two trees we've constructed so far will next be labeled by $\widehat p_{11}\widehat \BC_1+\widehat p_{4}\widehat \BC_4$ and $\widehat p_{12}\widehat \BC_1+\widehat p_{5}\widehat \BC_5$, respectively. The next step of our construction is to now check again for which $B$-nodes can be merged, and we will find (according to \eqref{eqnEx5}) that $\widehat\BC_6$ lies in the cone of $\widehat\BC_1$ and $\widehat\BC_4$ while $\widehat\BC_7$ lies within that of $\widehat\BC_1$ and $\widehat\BC_5$. Hence with the constraints,
\begin{eqnarray}\label{eqn3}
	\widehat p_{6}\widehat \BC_6&=&\widehat p_{11}\widehat \BC_1+\widehat p_{4}\widehat \BC_4\nonumber\\
	\widehat p_{7}\widehat \BC_7&=&\widehat p_{12}\widehat \BC_1+\widehat p_{5}\widehat \BC_5,
\end{eqnarray}
we can merge $\widehat q_6\widehat\AC_6\otimes\widehat p_6\widehat\BC_6$ and $\widehat q_7\widehat\AC_7\otimes\widehat p_7\widehat\BC_7$, one to each of the left-most $B$-nodes in part (a) of the figure, and those two $B$-nodes can then be relabeled in agreement with how they are labeled in part (b).

Next, attach new $A$-nodes to the left of each of these trees and label them as $\widehat q_{4}\widehat \AC_4+\widehat q_{6}\widehat \AC_6$ and $\widehat q_{5}\widehat \AC_5+\widehat q_{7}\widehat \AC_7$, respectively. When next checking for intersections of convex cones for $A$-nodes, we will find that these two nodes do indeed share an intersection. Therefore, merge them into one node, labeled by either of these sums, and attach a new $B$-node to the left, with label $\widehat p_6\widehat \BC_6+\widehat p_{7}\widehat \BC_7$. We obtain one additional constraint,
\begin{eqnarray}\label{eqn2500}
	\widehat q_4\widehat \AC_4+\widehat q_6\widehat \AC_6=\widehat q_{5}\widehat \AC_5+\widehat q_7\widehat \AC_7.
\end{eqnarray}

We now have a single tree in which every $\widehat\AC_j\otimes\widehat\BC_j$ appears at least once, so we can include two additional constraints that the two roots of this tree are $I_A,I_B$,
\begin{eqnarray}\label{eqn4000}
	I_A&=&\widehat q_4\widehat \AC_4+\widehat q_{6}\widehat \AC_6\nonumber\\
	I_B&=&\widehat p_6\widehat \BC_6+\widehat p_{7}\widehat \BC_7.
\end{eqnarray}
Thus, we have constructed a closed, double-rooted tree for this set of measurement outcomes. Equations \eqref{eqn1} through \eqref{eqn4000} constitute the complete set of constraints that must be solved to determine if this tree represents a valid LOCC protocol. One solution to these constraints is indicated by \eqref{eqnEx5}, which does indeed yield such a protocol. Notice that even though there are only seven outcomes defining the measurement, there are actually eight final outcomes of the protocol, an obvious consequence of the fact that $\widehat\AC_1\otimes\widehat\BC_1$ appears twice. 
\end{description}

\section{Conclusions}\label{conc}
To summarize, we have seen that every LOCC protocol corresponds to a tree graph with nodes labeled in a way that must satisfy (\ref{ACsum}) and (\ref{ACprop}), which constrain the way that any such tree may be constructed from a separable measurement. Given a separable measurement, then, we know how to construct an LOCC protocol if one exists. Furthermore, we see that by constructing all~\footnotemark[\value{footnote}] possible trees of depth $L$ working backward from the leaves, we will then also determine when no LOCC consisting of $L$ or fewer rounds exists for this measurement. We note, finally, that it is not difficult to generalize the construction to the case of more than two parties.

\textit{Acknowledgments} --- I am very grateful to Li Yu, Jonathan Walgate, Ronald de Wolf, Vlad Gheorghiu, and  Shiang-Yong Looi for helpful discussions, and to the KITP at the University of California, Santa Barbara, where this paper was begun. This work has been supported in part by a grant from the Research Corporation, and by the National Science Foundation through Grants No. PHY-0757251 and No. PHY-0551164.

\appendix
\section{Proof of main theorem}\label{sct3}
Here, we will see that the construction described in the main text produces an LOCC protocol for a given separable measurement corresponding to positive operators $\{\widehat\AC_j\otimes\widehat\BC_j\}$, whenever one exists. Every LOCC protocol can be described as a sequence of complete local measurements, each conditioned on previous outcomes. Each local measurement is a branching of possibilities to one of the outcomes of this measurement, and the collection of those outcomes that follow any given outcome of the previous measurement satisfies a completeness relation, as expressed by \eqref{closure}. This branching lends itself directly to representation as a tree graph (without closed loops), which we have illustrated in various figures. As has been described in the main text, the nodes of the tree can be labeled by the positive operators corresponding to the product of Kraus operators implemented by that party up to that point in the protocol, and we have shown that these positive operators must satisfy \eqref{ACsum} and \eqref{ACprop} at each and every node of the tree [\eqref{ACprop} is a direct consequence of \eqref{ACsum}, along with the branching structure of the tree]. Hence, if we can construct all trees (if any) that (\textit{i}) are compatible with these two equations at every node; (\textit{ii}) end with one of the $\widehat\AC_j\otimes\widehat\BC_j$ on each of the leaves; and (\textit{iii}) have each $\widehat\AC_j\otimes\widehat\BC_j$ appearing on at least one of the leaves; then we will have found all LOCC protocols for this separable measurement. Since every LOCC protocol corresponds to a tree, if there is no such tree, then there is no LOCC protocol.

We will not be able to construct all LOCC trees, however. The first reason is that we do not presently have an argument that our construction will always terminate on its own, so it is possible the algorithm of section~\ref{algorithm} could continue indefinitely. We therefore impose an upper limit $L$, which is in principle arbitrary, on the number of levels (depth) of the trees we construct, equivalent to restricting the number of rounds in the LOCC protocols that are considered.

To understand the second reason we will not construct all LOCC trees, even those of restricted depth, divide the entire collection of trees of depth no greater than $L$ into two classes. Class $\CC$ (for congruent) contains all such trees that include at least one node from which emerges (to the right) two or more sub-trees that are \emph{congruent}, by which we mean they are identical to each other apart from differing coefficients $\widehat q_{jk},\widehat p_{jk}$ (that is, differing $k$-indices) multiplying the operators on their leaves (see Fig.~\ref{fig:treeIdntclApndx} in appendix~\ref{ssct1} for an example). Since \emph{any} number of such congruent sub-trees can emerge from any given node, and since this number can be changed without altering the fact that the complete tree is a valid LOCC (again, see appendix~\ref{ssct1}), the number of trees in this class is infinite. Constructing all trees in this class is therefore a difficult, perhaps intractable, problem. The second class of trees is finite, and is the complement of the first within the entire set of LOCC trees of depth not exceeding $L$. This latter class, denoted $\DC$ (for distinct), contains all trees in which every node has \emph{only} distinct sub-trees emerging from it to the right, no two of which are congruent.

We will construct all trees in $\DC$, and then we will show in appendix~\ref{ssct1} that if no LOCC tree can be constructed in $\DC$, then there are no LOCC trees within $\CC$, either. Therefore, if our construction fails, then no LOCC protocol with $L$ or fewer rounds exists for this separable measurement.

We will next give a proof of our main theorem, that the construction will yield an LOCC protocol whenever one exists in $L$ or fewer rounds.

\subsection{The construction yields an LOCC tree whenever one exists}\label{ssct3a}
We here give a restatement of our main theorem.

\textbf{Main Theorem} (restatement)\label{thm1} \textit{The construction given in algorithmic form in section~\ref{algorithm}, when restricted to $L$ rounds, builds an LOCC tree for a given set of product operators $\{\widehat\AC_j\otimes\widehat\BC_j\}$ whenever an LOCC protocol in $L$ rounds exists for this separable measurement ($L$ is a finite, but otherwise arbitrary, integer).}

The main idea of the proof is to imagine that an (arbitrary) LOCC protocol is provided to us and then to show that the algorithm will construct an LOCC protocol for the same measurement. Since the algorithm starts from a known set of $\{\widehat\AC_j\otimes\widehat\BC_j\}$, we identify these operators by examining the leaves of the tree associated with the given protocol. Then, from this measurement and the known structure of its LOCC tree, the proof proceeds to show how the algorithm constructs that tree. Since the tree we started with was arbitrary, this proves that the algorithm will construct an LOCC protocol whenever one exists.

\noindent \emph{Proof}: Suppose an LOCC protocol exists, and consists of $L$ rounds; then there is an LOCC tree associated with that protocol. If this tree includes direct merging of congruent sub-trees, then according to lemma \ref{lem1} (proved below in appendix~\ref{ssct1}), there is also an LOCC tree for this measurement that has no such direct merging. By lemma~\ref{lem2} of appendix~\ref{sct2}, the latter tree yields an LOCC protocol. Therefore, we need only show that the construction builds the latter tree, the one without any direct merging of congruent sub-trees, so consider the latter (from here on referred to as the `original' tree) in the following. We may always assume that every branch has the same number of nodes along it by extending shorter ones with additional nodes along a single line (no further branching), each such node representing a round where that party did nothing (performed the identity operator). For the same reason, we can (arbitrarily) choose to have every leaf be an $A$-node emerging from a $B$-node on its left.

Imagine cutting all the edges joining the latter $B$-nodes to the $A$-nodes on their left (not to the leaf $A$-nodes, which are on their right). This separates all those $B$-nodes from each other, leaving multiple `two-level' trees each with a single $B$-node and one or more (leaf) $A$-nodes attached to it. No two of those $A$-nodes emerging from a single, given $B$-node will be the same $\widehat\AC_j$, as this would entail a direct merging of congruent sub-trees: two $\widehat\AC_j\otimes\widehat\BC_j$ with the same $j$ will not be merged directly to each other in this tree. Each of these individual trees is a merging of several different $\widehat\AC_j\otimes\widehat\BC_j$ together at the $B$-node, where those $B$-nodes must all be proportional to each other. That is,
\begin{align}\label{eqn1000}
\BC_{i_{\!L\!-\!1}}^{(\SC_{\!L\!-\!1})}=p_{jk}\widehat\BC_{j}=p_{j^\prime k^\prime}\widehat\BC_{j^\prime}=\cdots,
\end{align}
where $\BC_{i_{\!L\!-\!1}}^{(\SC_{\!L\!-\!1})}$ is the label on that particular $B$-node in the original tree. [Note that we have drawn a distinction between the coefficients $q_{jk}, p_{jk}$ from the original tree, which are \textit{known} (by assumption the original tree is given in advance, so is fully known), and the $\widehat q_{jk}, \widehat p_{jk}$ to be used in our construction, which are as yet \textit{unknown}.] Since $\widehat\BC_{j}, \widehat\BC_{j^\prime}, \cdots$ are all proportional to each other, then at the first pass through step \ref{enum1} of the algorithm our construction will build each of the two-level trees that result from this cutting of the original tree, introducing unknown coefficients $\widehat q_{jk}, \widehat p_{jk}$ on the $A$ and $B$ nodes. The algorithm also adds a new $A$ node on the left to each of these, and labels them with sums of those $\widehat q_{jk}\widehat\AC_j$. [The cut we made in the original tree may have produced multiple copies of a given (sub-)tree. To this point in the construction, we will only have built a single copy because we will not yet have (fore)seen the need for more than one, but the additional copies will arise later. As discussed above, the multiple copies will not be directly merged to each other in the original tree, so they must merge to other sub-trees before the resulting composite sub-trees merge to each other, where those other sub-trees differ from one copy of this sub-tree to another. When multiple trees that can be merged with a given (fixed) one are present at some stage in our construction, then the algorithm makes the necessary multiple copies of the fixed one so that it can be merged with each of those others. This is done by identifying multiple classes and/or multiple subsets of those classes within which that fixed tree is included. Even at this first stage, this will generally already have occurred, if any single $\widehat\AC_j\otimes\widehat\BC_j$ appears more than once on the leaves of the original tree.]

Returning to the entire tree, cut edges one level further into the tree from the leaves, isolating some number of three-level trees (rather than the two-level ones obtained by the cut made in the previous paragraph), all having $A$-nodes on their left with one or more $B$-nodes emerging to the right. The left-most $A$-node in each of these trees must satisfy \eqref{ACsum}, which says that it is equal to a sum of the $q_{jk}\widehat\AC_j$ that label all the leaf $A$-nodes that emerge from any one of the $B$-nodes emerging from that left-most $A$-node. If there are multiple such $B$-nodes emerging, then the multiple sums of those individual sets of $q_{jk}\widehat\AC_j$ necessarily satisfy \eqref{ACprop}, since by assumption the original tree is a valid LOCC. But satisfying \eqref{ACprop} implies that the convex cones generated by those sets of $\widehat\AC_j$ are mutually intersecting, and by the previous paragraph, those same sets of $\widehat\AC_j$ appear in the sums labeling the new nodes that have just been added in our construction. Therefore in our construction, a tree will appear that merges those same $A$-nodes as are merged at this level in the original tree (with $\widehat q_{jk}, \widehat p_{jk}$ in place of the original $q_{jk}, p_{jk}$). 
Therefore, our construction will build all these three-level trees from the two-level ones previously obtained.

Now we can use an inductive argument. Assume that our construction builds all $l$-level trees that appear in the original tree and labels them with the $\widehat\AC_j,\widehat\BC_j$ in exactly the same way as they are labeled in that tree, apart from replacing $q_{jk}\rightarrow\widehat q_{jk}$ and $p_{jk}\rightarrow\widehat p_{jk}$. We have just shown this holds for $l=2,3$, so assuming it holds for all levels up to $l$, if we can show that it then also holds for $l+1$, the proof will be complete. Since all the $l$-level trees are labeled by sums involving the same set of $\widehat\AC_j,\widehat\BC_j$ as in the original tree, we know that when we look for intersections of convex cones generated by these sets, we will always find intersections that will lead us to merge these $l$-level trees to create every one of the $(l+1)$-level trees
that appear from that cut of the original tree (the cut that produces $(l+1)$-level trees). The reason is that these intersections are simply solutions to equations of the form \eqref{ACprop}, and we know from the original tree that at least one solution will always exist, since $\widehat q_{jk}=q_{jk},\widehat p_{jk}=p_{jk}$ is such a solution, and this completes the proof.\hspace{\stretch{1}}$\blacksquare$

Note that for a given set of $\{\widehat\AC_j\otimes\widehat\BC_j\}$, if an LOCC protocol exists in any finite number $L^\prime$ of rounds, then by setting $L=L^\prime$ in the above arguments, we have the immediate corollary that our construction will build an LOCC protocol whenever one exists in \emph{any} finite number of rounds.

Let us illustrate the procedure of the proof by an example. Consider the first cut isolating two-level trees that lie at the leaf end of the original tree, nodes labeled by $\AC_{i_{\!L}}^{(\SC_{\!L})}$ and $\BC_{i_{\!L\!-\!1}}^{(\SC_{\!L\!-\!1})}$, respectively. Suppose node $\BC_{1}^{(\SC_{\!L\!-\!1})}$ branches to three $\AC_{i_{\!L}}^{(\SC_{\!L})}$ ($\SC_{\!L}=\{\SC_{\!L\!-\!1},1\}$) with $i_{\!L}=1,2,3$. Since this LOCC implements the $\{\widehat\AC_j\otimes\widehat\BC_j\}$, it must be that (for some ordering of the latter and for some set of coefficients $q_{j1}, p_{j1}$),
\begin{align}\label{eqn10}
\BC_{1}^{(\SC_{\!L\!-\!1})}=p_{11}\widehat\BC_1=p_{21}\widehat\BC_2=p_{31}\widehat\BC_3,
\end{align}
and
\begin{align}\label{eqn11}
\AC_{1}^{(\SC_{\!L})}=q_{11}\widehat\AC_1\notag\\
\AC_{2}^{(\SC_{\!L})}=q_{21}\widehat\AC_2\notag\\
\AC_{3}^{(\SC_{\!L})}=q_{31}\widehat\AC_3.
\end{align}

What this means is that $\widehat\BC_1,\widehat\BC_2,\widehat\BC_3$ are all proportional to each other (each generates a one-dimensional convex cone and all three of these cones share a mutual intersection as they are identical to each other), from which it immediately follows that the three two-node trees, $\widehat q_{j1}\widehat\AC_j\otimes \widehat p_{j1}\widehat\BC_j,~j=1,2,3$, will be merged at the $B$-nodes in our construction forming a tree having the exact same structure as the two-level tree in which $\BC_{1}^{(\SC_{\!L\!-\!1})}$ branches to the three nodes $\AC_{i_{\!L}}^{(\SC_{\!L})},~i_{\!L}=1,2,3$, in the original tree. Not only is the structure exactly the same, but in fact the two trees are identical in every way except that the one obtained from our construction has coefficients that are as yet unknown. Although this is a specific example of a $B$-node branching to three $A$-nodes, it should be clear that the ideas are completely general, and for each and every final two-level sub-tree arising from this cut --- $\BC_{i_{\!L\!-\!1}}^{(\SC_{\!L\!-\!1}^\prime)}$ with (possibly) multiple edges emerging to $A$-nodes $\AC_{i_{\!L}}^{(\SC_{\!L}^\prime)}$ on the leaves --- a tree will be formed in our construction that is identical to it apart from the unknown coefficients.

The next thing is to consider the cut done one level further into the tree from the leaves, isolating sets of three-level sub-trees. In our example, if node $\BC_{1}^{(\SC_{\!L\!-\!1})}$ emerges from node $\AC_1^{(\SC_{\!L\!-\!2})}$, then according to \eqref{ACsum}, it must be that
\begin{align}\label{eqn111}
\AC_1^{(\SC_{\!L\!-\!2})}=\AC_{1}^{(\SC_{\!L})}+\AC_{2}^{(\SC_{\!L})}+\AC_{3}^{(\SC_{\!L})}=q_{11}\widehat\AC_1+q_{21}\widehat\AC_2+q_{31}\widehat\AC_3.
\end{align}
Recall that in our construction, an $A$-node is attached at the left of any merged $B$-node. In the specific example we are here discussing, $\widehat\BC_1,\widehat\BC_2,\widehat\BC_3$ are all merged into a single node, and the newly attached $A$-node will be labeled in our construction as $\widehat q_{11}\widehat\AC_1+\widehat q_{21}\widehat\AC_2+\widehat q_{31}\widehat\AC_3$ (for the choice $\widehat q_{j1} = q_{j1}$, this is indeed equal to $\AC_1^{(\SC_{\!L\!-\!2})}$, demonstrating that at least to this point in the construction, there exists a solution for the coefficients).

Suppose node $\BC_{2}^{(\SC_{\!L\!-\!1})}$ also emerges from node $\AC_1^{(\SC_{\!L\!-\!2})}$ and emerging from that $\BC_{2}^{(\SC_{\!L\!-\!1})}$ node are two $\AC_{i_{\!L}}^{(\SC_{\!L}^\prime)},~i_L=1,2$ with $\SC_L^\prime=\{\SC_{L\!-\!1},2\}$. Then, again by \eqref{ACsum}, it must be that $\AC_{1}^{(\SC_{\!L}^\prime)}+\AC_{2}^{(\SC_{\!L}^\prime)}=\AC_1^{(\SC_{\!L\!-\!2})}$. By the same argument as was used above, we will have
\begin{align}\label{eqn12}
\BC_{2}^{(\SC_{\!L\!-\!1})}=p_{41}\widehat\BC_4=p_{51}\widehat\BC_5,
\end{align}
and
\begin{align}\label{eqn13}
\AC_{1}^{(\SC_{\!L}^\prime)}=q_{41}\widehat\AC_4\notag\\
\AC_{2}^{(\SC_{\!L}^\prime)}=q_{51}\widehat\AC_5
\end{align}
so that
\begin{align}\label{eqn14}
\AC_1^{(\SC_{\!L\!-\!2})}=q_{11}\widehat\AC_1+q_{21}\widehat\AC_2+q_{31}\widehat\AC_3=q_{41}\widehat\AC_4+q_{51}\widehat\AC_5,
\end{align}
and I have assumed for illustration purposes that $j=1,2,3$ are not repeated in the latter set of two $\widehat\AC_j\otimes\widehat\BC_j$, though that could certainly also occur. This demonstrates that the two convex cones generated by $\widehat\AC_1,\widehat\AC_2,\widehat\AC_3$ and by $\widehat\AC_4,\widehat\AC_5$ intersect with each other (and furthermore, $\AC_1^{(\SC_{\!L\!-\!2})}$ lies within that intersection). Therefore, at the second step of our construction, a tree will be created including all these nodes and having the same structure as the corresponding sub-tree within the original tree: an $A$-node branching to these two $B$-nodes, one of which branches to three $A$-nodes and the other to two. Again we can see that even though this is only a specific example the ideas can be applied quite generally, and every sub-tree within the original tree having a depth of three nodes will also be created in our construction. Furthermore, just as $\AC_1^{(\SC_{\!L\!-\!2})}$ lies within that intersection of the two convex cones mentioned above, all trees created in our construction to this level will have left-most nodes labeled in a way that the corresponding $\AC_{i_{\!L\!-\!2}}^{(\SC_{\!L\!-\!2})}$ will lie in the convex cone generated by the sum of the $\widehat\AC_j$ that labels the node in that tree we've created.

This sort of analysis can be continued back to the roots of the tree, and by similar arguments one will find that one of the trees created by our construction has a structure identical to the entire original tree. In addition, all nodes of this tree are labeled by positive linear combinations of the $\widehat\AC_j$ (or $\widehat\BC_j$), and the $\AC_{i_{n}}^{(\SC_{n})}$ (or $\BC_{i_m}^{(\SC_{m})}$) on a given node of that original tree will always lie within the convex cone generated by the $\widehat\AC_j$ (or $\widehat\BC_j$) labeling the corresponding node in this tree, precisely as was found above for the final few levels of the tree. Therefore, there will always be at least one solution to the constraints on the $\widehat q_{jk},\widehat p_{jk}$, that solution being $\widehat q_{jk}=q_{jk}$ and $\widehat p_{jk}=p_{jk}$, which yields precisely the original LOCC tree. Thus, we see that given any set $\{\widehat\AC_j\otimes\widehat\BC_j\}$ for which there exists an LOCC protocol, our construction always builds an LOCC tree, providing an associated LOCC protocol.

\section{LOCC trees and LOCC protocols}\label{sct2}
In this section, we show that there is an equivalence between LOCC trees and LOCC protocols, which will be useful in that it tells us that not only does every LOCC protocol correspond to such a tree, but also that if we can construct a tree of this type, then we can also construct an LOCC protocol.
\begin{lem2}\label{lem2}
A double-rooted tree represents an LOCC protocol if its roots are labeled by $I_A,I_B$ and every one of its nodes (i) has a single edge entering it from the left (excepting the left-most root); and (ii) satisfies \eqref{ACsum} for every branch emerging from it toward the right.
\end{lem2}
\noindent The reverse implication, that every LOCC protocol corresponds to such a tree was shown in the main text. Therefore, this lemma completes the demonstration of an equivalence between these trees and LOCC protocols.

\noindent \emph{Proof}: We will construct an LOCC protocol from any tree satisfying the conditions of the lemma. Each (non-root) node of the tree is labeled by $\AC_{i_n}^{(\SC_{n})}$ or $\BC_{i_n}^{(\SC_{n})}$, which satisfy \eqref{ACsum}, reproduced here for convenience,
\begin{eqnarray}\label{eqn100}
	\sum_{i_n}\AC^{(\SC_n)}_{i_{n}} = \AC^{(\SC_{n\!-\!2})}_{i_{n\!-\!2}},
\end{eqnarray}
and similarly for Bob's operators. The phrase `for every branch emerging from it toward the right' in (\textit{ii}) of the statement of the lemma should be understood in \eqref{eqn100} to mean `for every $i_{n-1}$', as each of those branches from node $(\SC_{n\!-\!2},i_{n\!-\!2})$ are labeled by one of the integers, $i_{n-1}$.

Suppose the first nodes to the right of the roots are $A$-nodes (the arguments are the same if they are $B$-nodes). Define Kraus operators 
\begin{eqnarray}\label{eqn101}
	A_{i_1}=\sqrt{\AC_{i_1}},
\end{eqnarray}
and choose the unique positive square root. Then, we have that
\begin{eqnarray}\label{eqn102}
	\sum_{i_1}A_{i_1}^\dag A_{i_1}=\sum_{i_1}\AC_{i_{1}} = I_A,
\end{eqnarray}
which follows from \eqref{eqn100} with $n=1$ along with the fact that the node that is two steps to the left of the $n=1$ nodes is the left-most root, equal to $I_A$. Therefore, this set of Kraus operators is a complete measurement, and we choose it as Alice's first one. By the same argument the Kraus operators
\begin{eqnarray}\label{eqn103}
	B^{(i_1)}_{i_2}=\sqrt{\BC^{(i_1)}_{i_2}}
\end{eqnarray}
for each fixed $i_1$, are a complete first measurement for Bob, conditioned on the outcome of Alice's first measurement.

Let us now show how to choose each of Alice's subsequent measurements; Bob's can be chosen by the same approach. For Alice's second measurement, choose
\begin{eqnarray}\label{eqn104}
	A^{(\SC_3)}_{i_3}=U^{(\SC_3)}_{i_3}\sqrt{\AC^{(\SC_3)}_{i_3}}~A_{i_1}^{-1},
\end{eqnarray}
where $\SC_3=(i_1,i_2)$, we choose unitary $U^{(\SC_3)}_{i_3}$ for the convenience of having $A^{(\SC_3)}_{i_3}$ be a positive operator (we will mean `positive semidefinite' whenever we write `positive'), and $A_{i_1}^{-1}$ should be understood to be the inverse on its support.  Then $A_{i_1}^{-1}A_{i_1}=P_{i_1}$, where $P_{i_1}$ is the projector onto the support of $A_{i_1}$ and is also equal to $A_{i_1}A_{i_1}^{-1}$ since $A_{i_1}$ is a positive operator. Include one additional Kraus operator equal to $I_A-P_{i_1}$. Now,
\begin{eqnarray}\label{eqn105}
	\sum_{i_1}A^{(\SC_3)\dag}_{i_3} A^{(\SC_3)}_{i_3}=\sum_{i_3}\left(A_{i_1}^\dag\right)^{-1}\AC^{(\SC_3)}_{i_3}A_{i_1}^{-1} = \left(\sqrt{\AC_{i_1}}\right)^{-1} \AC_{i_1} \left(\sqrt{\AC_{i_1}}\right)^{-1} = P_{i_1},
\end{eqnarray}
where the second equality follows from \eqref{eqn100}. Adding in that last Kraus operator shows this is a complete measurement. The overall Kraus operator (product of Kraus operators implemented by Alice up to any given point in the protocol) corresponding to this added one is $(I_A-P_{i_1})A_{i_1}=0$, which is why it need not appear as a node in our tree. [If one wishes, an additional branch and node can be added for this `outcome' of the measurement, with no further nodes emerging from it. Since it vanishes identically, this is really unnecessary.]

Assume we have managed to construct a complete set of (positive) Kraus operators for each of Alice's local measurements up to level $m$. If we can then construct a complete set of Kraus operators for her next measurement ($m+2$), we will have shown by induction that the tree yields an LOCC protocol, completing the proof. Thus by assumption, we have positive $A^{(\SC_m)}_{i_m}$ such that
\begin{eqnarray}\label{eqn106}
	\sum_{i_m}A^{(\SC_m)\dag}_{i_m} A^{(\SC_m)}_{i_m}=\sum_{i_m}\AC^{(\SC_{m})}_{i_{m}}=I_A,
\end{eqnarray}
where we have included an extra Kraus operator like the one introduced in the previous paragraph. Define
\begin{eqnarray}\label{eqn107}
	A^{(\SC_{\!m\!+\!2})}_{i_{\!m\!+\!2}}=U^{(\SC_{\!m\!+\!2})}_{i_{\!m\!+\!2}}\sqrt{\AC^{(\SC_{\!m\!+\!2})}_{i_{\!m\!+\!2}}}~\left(A^{(\SC_m)}_{i_m}\right)^{-1},
\end{eqnarray}
choosing $U^{(\SC_{\!m\!+\!2})}_{i_{\!m\!+\!2}}$ so that each of these is positive. Then,
\begin{eqnarray}\label{eqn108}
	\sum_{i_{\!m\!+\!2}}A^{(\SC_{\!m\!+\!2})\dag}_{i_{\!m\!+\!2}}A^{(\SC_{\!m\!+\!2})}_{i_{\!m\!+\!2}}=\sum_{i_{\!m\!+\!2}}\left(A^{(\SC_m)\dag}_{i_m}\right)^{-1}\AC^{(\SC_{\!m\!+\!2})}_{i_{\!m\!+\!2}}\left(A^{(\SC_m)}_{i_m}\right)^{-1}=\left(A^{(\SC_m)\dag}_{i_m}\right)^{-1}\AC^{(\SC_{m})}_{i_{m}}\left(A^{(\SC_m)}_{i_m}\right)^{-1}=P^{(\SC_m)}_{i_m},
\end{eqnarray}
and after including that one additional Kraus operator equal to $I_A-P^{(\SC_m)}_{i_m}$ with $P^{(\SC_m)}_{i_m}=A^{(\SC_m)}_{i_m}\left(A^{(\SC_m)}_{i_m}\right)^{-1}=\left(A^{(\SC_m)}_{i_m}\right)^{-1}A^{(\SC_m)}_{i_m}$, we see that this is a complete set for Alice's next measurement. Note once again that $(I_A-P^{(\SC_m)}_{i_m})A^{(\SC_m)}_{i_m}=0$, so this extra Kraus operator has zero probability of occcurring in the protocol, completing the proof. \hspace{\stretch{1}}$\blacksquare$

If one wishes to obtain a specific set of Kraus operators at the end of the protocol (levels $m=L-1,L$), and if these Kraus operators are compatible with the positive operators $\AC_{i_m}^{(\SC_m)},\BC_{i_m}^{(\SC_m)}$ (for those same values of $m$), this can always be done by adjusting the $U^{(\SC_{m})}_{i_{m}}$ at these levels.

\section{Merging congruent sub-trees directly to one another}\label{ssct1}
In this section, we provide the final piece of the puzzle by proving the following result which was assumed in the proof of theorem~\ref{thm1}.
\begin{lem1}\label{lem1}
If an LOCC tree exists for a given set of product operators $\{\widehat\AC_j\otimes\widehat\BC_j\}$ and includes direct merging of congruent sub-trees, then there also exists an LOCC tree for the same set of operators that does not include any such direct merging.
\end{lem1}
\noindent The reason it is not necessary to consider trees with congruent sub-trees merged directly to one another is that those congruent sub-trees can easily be combined into a single sub-tree. Note that being congruent, every node in each of these sub-trees has a counterpart in the other sub-trees, where all these counterparts are labeled by linear combinations of the exact same set of operators, the only difference being that the coefficients, $\widehat q_{jk},\widehat p_{jk}$, are different. This follows from \eqref{ACsum} and the fact that the sub-trees are congruent, which implies that the set of final outcomes (and the number of copies of each) at the leaves of these sub-trees are the same.

\begin{figure}
\includegraphics{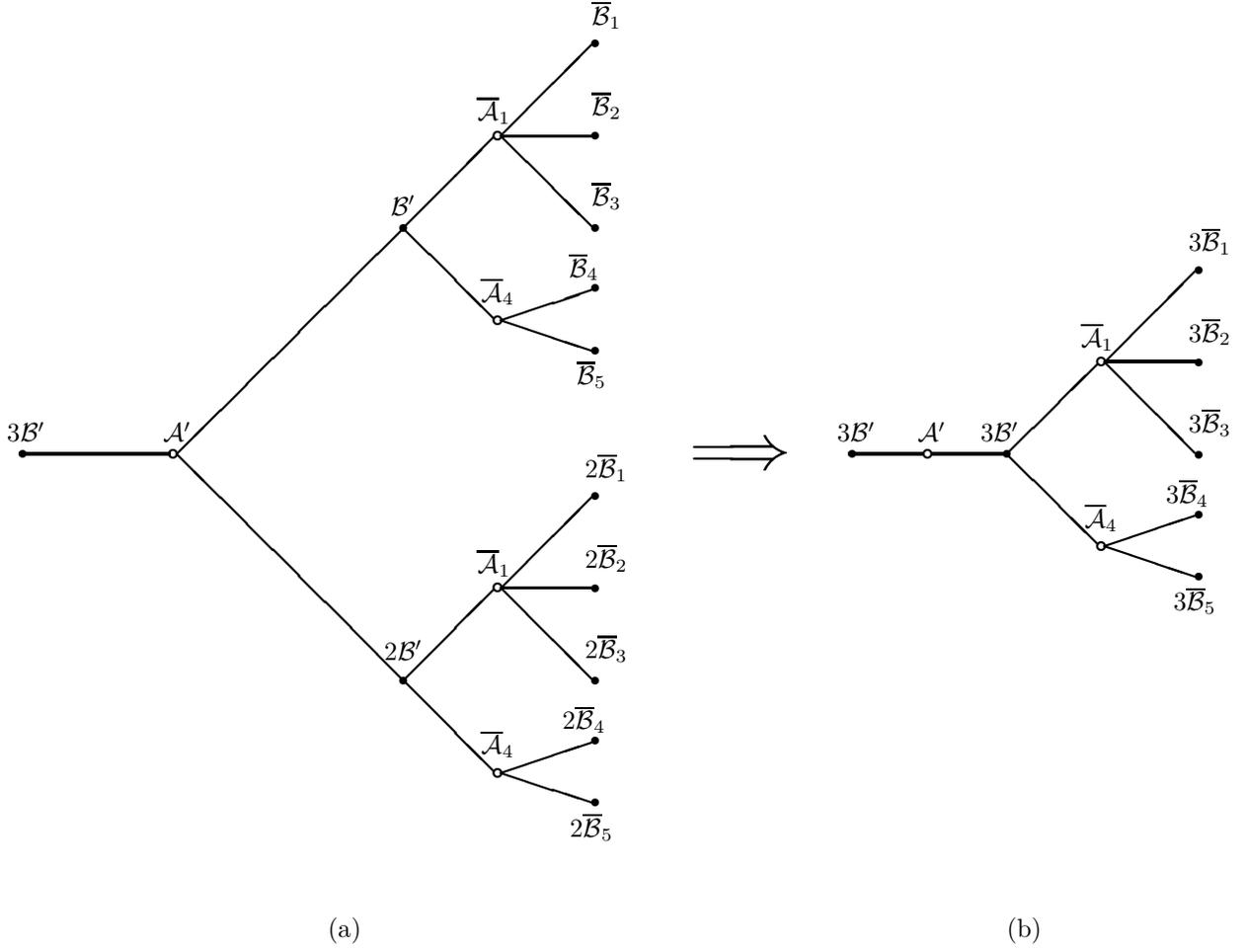}
\caption{\label{fig:treeIdntclApndx}Illustrating why it is not necessary to consider merging multiple copies of congruent sub-trees. To simplify the notation in the figure, all coefficients ($\widehat q_{jk},\widehat p_{jk}$) have been taken to be integers for illustration purposes. We have that $\AC^\prime=\widehat\AC_1+\widehat\AC_4$ and $\BC^\prime=\widehat\BC_1+\widehat\BC_2+\widehat\BC_3=\widehat\BC_4+\widehat\BC_5$. Any valid LOCC tree in which sub-tree (a) on the left appears can be replaced by a valid LOCC tree that has this sub-tree replaced by the simpler one (b) on the right. The reason is that the left-most nodes (one $A$ and one $B$) are the same in these two sub-trees, so whatever else (a) connects to can just as well be connected to (b).}
\end{figure}

Before proceeding to the proof of lemma \ref{lem1}, let us first illustrate how, from a tree that has several congruent sub-trees that are merged directly, we can construct a tree that has only one of those sub-trees. The node at which they are merged corresponds to one of the parties, say $A$. Keep everything in the original tree unchanged except for these congruent sub-trees. Then,
\begin{enumerate}
  \item Erase all but one of those congruent sub-trees in their entirety;
  \item Leave all the $A$-node labels unchanged in the remaining sub-tree (because it is an $A$-node where the congruent sub-trees are merged);
  \item Replace each $B$-node label in the remaining sub-tree by a sum --- over all counterpart nodes for that particular node, including the one in the sub-tree that remains --- of the linear combination of operators that label those counterparts (this sums the $\widehat p_{jk}$'s appearing on all those counterpart nodes, for each fixed $j$).
\end{enumerate}
This is illustrated in Fig.~\ref{fig:treeIdntclApndx}. If the original tree is valid for LOCC, then so is the new one. The reason this works is that the left-most nodes (one $A$ and one $B$) are the same in the two sub-trees (a) and (b) in the figure, so whatever else sub-tree (a) connects to can just as well be connected to (b). If (a) is connected to other sub-trees at the $3\BC^\prime$ node (note that this node is not part of the congruent sub-trees), then those other sub-trees can also be connected to the left-most $3\BC^\prime$ node in (b). Everything else in the entire tree will remain unchanged. If, on the other hand, other sub-trees are connected to the $\AC^\prime$ node in (a), they can also be connected to the $\AC^\prime$ node in (b), though then the label $3\BC^\prime$ on the left-most node in both (a) and (b) would need to be altered, but this alteration will be the same in (b) as it is in (a). It should be noted that, while the new tree yields a valid LOCC protocol if the original one did, the new tree may correspond to a different set of weights $\widehat r_j=\sum_k\widehat q_{jk}\widehat p_{jk}$ as opposed to those in the original tree. This may be important to Alice and Bob depending on the context in which they wish to use the LOCC protocol. We leave for future work the question of when it will be possible to construct a larger tree from a smaller one by introducing direct merging of congruent nodes, with the aim of obtaining a specified set of weights.

Now we turn to the proof of the lemma.

\noindent\emph{Proof of lemma~\ref{lem1}}: If we can combine two congruent sub-trees into one, then we can combine any number of congruent sub-trees into one simply by combining them two at a time. Therefore, consider that sub-trees $T_1$ and $T_2$ are combined to become $\widehat{T}$ according to the prescription described above: erase $T_2$ and in $T_1$ leave the $A$-nodes unchanged and replace each $B$-node by the sum of that node with its counterpart from $T_2$. Since the $A$-nodes are unchanged throughout the entire tree, then the sums in \eqref{ACsum} and \eqref{ACprop} for the $\AC_{i_m}^{(\SC_m)}$ are also unchanged, so these required conditions will still be satisfied, assuming they were satisfied in the original tree. Thus, we need only demonstrate that these conditions will also still be satisfied for the $B$-nodes.

Label each $B$-node in $T_1$ as $\BC_{i_m}^{(\SC_m)}$, the corresponding node in $T_2$ as $\BC_{i_m}^{(\SC_{\!m}^\prime)}$, and that in $\widehat{T}$ as $\widehat{\BC}_{i_m}^{(\SC_m)}$. The indices denoting position in $T_1$ and in $\widehat{T}$ are exactly the same ($\SC_m,i_m$) because these two sub-trees have the exact same structure and lie in the same position within their respective overall trees. The indices denoting position within $T_2$ are not quite the same as those, differing only at the node where $T_1$ and $T_2$ are merged to each other, so that $\SC_{\!m}^\prime$ differs from $\SC_m$ only in the index corresponding to the different branch the two follow from that particular node. By the procedure described above, we have
\begin{align}\label{eqn20}
\widehat{\BC}_{i_m}^{(\SC_m)}=\BC_{i_m}^{(\SC_m)}+\BC_{i_m}^{(\SC_{\!m}^\prime)}
\end{align}
for each $m$ that labels a node in $T_1$ (and therefore, also in $T_2$), and by \eqref{ACsum}, we know that
\begin{align}\label{eqn21}
\BC_{i_m}^{(\SC_{m})}&=\sum_{i_{\!m\!+\!2}}\BC_{i_{\!m\!+\!2}}^{(\SC_{\!m\!+\!2})},\notag\\
\BC_{i_m}^{(\SC_{\!m}^\prime)}&=\sum_{i_{\!m\!+\!2}}\BC_{i_{\!m\!+\!2}}^{(\SC_{\!m\!+\!2}^\prime)}.
\end{align}
Adding the last two equations together, we obtain
\begin{align}\label{eqn10000}
\widehat{\BC}_{i_m}^{(\SC_{\!m})}=\sum_{i_{\!m\!+\!2}}\left(\BC_{i_{\!m\!+\!2}}^{(\SC_{\!m\!+\!2})}+\BC_{i_{\!m\!+\!2}}^{(\SC_{\!m\!+\!2}^\prime)}\right)=\sum_{i_{\!m\!+\!2}}\widehat{\BC}_{i_{\!m\!+\!2}}^{(\SC_{\!m\!+\!2})},
\end{align}
showing that within $\widehat{T}$, \eqref{ACsum} is satisfied at every $B$-node.

The rest of the original tree outside $\widehat{T}$ has been unaltered by this procedure for merging $T_1$ with $T_2$. We need to check that \eqref{ACsum} will continue to be satisfied at every $B$-node outside of $\widehat{T}$ in the newly formed tree. Suppose the $A$-node at which $T_1$ and $T_2$ were merged is $\AC_{i_{\!M\!+\!1}}^{(\SC_{\!M\!+\!1})}$. By \eqref{ACsum}, the $B$-node $\BC_{i_M}^{(\SC_{\!M})}$ from which $\AC_{i_{\!M\!+\!1}}^{(\SC_{\!M\!+\!1})}$ emerges is, in the original tree, equal to
\begin{align}\label{eqn101}
\BC_{i_M}^{(\SC_{\!M})}=\sum_{i_{\!M\!+\!2}}\left(\BC_{i_{\!M\!+\!2}}^{(\SC_{\!M\!+\!2})}+\BC_{i_{\!M\!+\!2}}^{(\SC_{\!M\!+\!2}^\prime)}\right)+ \Delta,
\end{align}
where the terms in the sum are from $T_1$ and $T_2$, respectively. Quantity $\Delta$ includes everything that contributes from outside $T_1$ and $T_2$, and is unchanged in the new tree, in which $\BC_{i_M}^{(\SC_{\!M})}$ has also not changed from what it was in the original tree (it is to the left of $\AC_{i_{\!M\!+\!1}}^{(\SC_{\!M\!+\!1})}$ and therefore not a part of $T_1,T_2$). Therefore, $\BC_{i_M}^{(\SC_{\!M})}$ remains equal to \eqref{eqn101}, whereas by \eqref{ACsum}, it should be equal to
\begin{align}\label{eqn102}
\BC_{i_M}^{(\SC_{\!M})}=\sum_{i_{\!M\!+\!2}}\widehat{\BC}_{i_{\!M\!+\!2}}^{(\SC_{\!M\!+\!2})}+ \Delta,
\end{align}
in the new tree, which it clearly is.

We now argue that all other $B$-nodes in the new tree also satisfy \eqref{ACsum}. Those $\BC_{i_m}^{(\SC_{\!m})}$ that are not in $\widehat{T}$ and not downstream (to the left) of it, are unchanged from the original tree, as are their upstream $B$-neighbors $\BC_{i_{\!m\!+\!2}}^{(\SC_{\!m\!+\!2})}$, so it holds for these nodes. Those nodes downstream from $\widehat{T}$ are $\BC_{i_M}^{(\SC_{\!M})}$ and those downstream from it, such as $\BC_{i_{\!M\!-\!2}}^{(\SC_{\!M\!-\!2})}$, $\BC_{i_{\!M\!-\!4}}^{(\SC_{\!M\!-\!4})}$, etc. The latter are each, by \eqref{ACsum}, equal to sums of the $B$-nodes that lie immediately to the right in the original tree, and those $B$-nodes to the right are the same in the new tree as they were in the original one, so they still satisfy \eqref{ACsum}, which is therefore seen to be satisfied at every node in the new tree. Since the roots are also unchanged in the new tree, they remain equal to $I_A,I_B$ (even if these roots are the previously mentioned nodes $\AC_{i_{\!M\!+\!1}}^{(\SC_{\!M\!+\!1})}$ and $\BC_{i_M}^{(\SC_{\!M})}$). Thus, by lemma~\ref{lem2}, this tree is LOCC and yields a valid LOCC protocol, completing the proof.\hspace{\stretch{1}}$\blacksquare$

\section{Complexity of the construction}\label{sct4}

As indicated in the algorithm presented above for our construction of LOCC trees, the maximum number of trees generated at the $l^{\textrm{th}}$ pass through the algorithm is $N_l=2^{N_{l-1}}-1$, where $N_{l-1}$ is the maximum number of trees that could have been created at the previous pass through. Therefore, the complexity of this method of searching for an LOCC tree could be enormous, apparently as much as ``multiply-exponential", the space required to store all the trees as one proceeds being ${\cal O}\left(2^{N_L}\right)$ for $L$ rounds. Nonetheless, this is a finite upper bound on the number of trees that can be generated in constructing a protocol having no more than $L$ rounds. It should also be pointed out that this complexity is very much a `worst-case scenario', and is likely to be a rather loose upper bound. We believe, though have no proof, that almost all cases (and perhaps even \emph{all} cases) will require much less in the way of computational resources than this bound represents.

The total number of final outcomes (leaves) in any protocol constructed in this way is finite, but potentially quite large. For example, one might generate $N_l$ different trees after $l$ passes through the algorithm, and each of these large number of trees will generally have a large number of leaves on it. Then one could imagine at the next pass it might be possible to combine all these trees into a single tree, so that all the leaves of those trees become leaves of this one combined tree. Clearly, the total number of final outcomes would then be enormous. There is nonetheless a finite upper bound, and one can always devise a protocol that has more final outcomes than that bound. Does this then mean that there exist measurements for which an LOCC protocol exists, but for which our construction fails to find one? The answer is no for the reason that if one allows direct merging of congruent sub-trees (see the following section), then one can end up with any number of final outcomes in such a tree. Our construction will not build this tree, but by lemma \ref{lem1}, it will still yield an LOCC protocol for the same measurement, just one that has a smaller number of final outcomes.  The direct implication is that in order for a protocol to have a number of final outcomes exceeding that finite upper bound, the corresponding tree \emph{must} include direct merging of congruent sub-trees.


\begin{thebibliography}{10}
\bibitem{superNote}
The difference between our definition of a separable measurement and the more
  general concept of a separable operation is that in the latter, represented
  by $S(\rho)=\sum_j(\widehat A_j\otimes\widehat B_j)\rho(\widehat
  A_j^\dagger\otimes\widehat B_j^\dagger)$, the choice of Kraus operators is
  not uniquely determined.
\bibitem{ACNote}
This method will label every node by sums of $\widehat \AC_j$ or $\widehat
  \BC_j$. When a complete tree is constructed, one can then (if one wishes,
  though there is no particular need to do so) directly read off the values of
  the $\AC_{i_n}^{(\SC_n)}$ and $\BC_{i_n}^{(\SC_n)}$.
\bibitem{trivial}
Every pair of convex cones in a given space intersect at the origin. However,
  since such `trivial' intersections are not of interest to us here, the term
  `intersection' should be understood to mean `non-trivial intersection'
  throughout this paper.
\end{thebibliography}

\end{document}